\documentclass[cernpreprint, atlasdraft=true, texlive=2016,UKenglish, texmf]{atlasdoc}
 
\usepackage{atlaspackage}
\usepackage{atlasbiblatex}
 
\usepackage{atlasphysics}
\usepackage{atlasbsm}
\usepackage{atlasprocess}
\usepackage{atlasjetetmiss}

\graphicspath{{logos/}{figures/}}
 
\usepackage{ANA-SUSY-2018-19-PAPER-defs}
\usepackage{graphicx}
\usepackage{pdflscape}
\usepackage{braket}
\usepackage{multirow}

\AtlasTitle{Search for long-lived charginos based on a disappearing-track signature using 136 fb$^{-1}$ of $pp$ collisions at $\sqrt{s}$~=~13~TeV with the ATLAS detector}
 
\AtlasVersion{3.2}
\AtlasJournal{EPJC}
\AtlasJournalRef{Eur. Phys. J. C 82 (2022) 606}
\AtlasDOI{10.1140/epjc/s10052-022-10489-5}
\AtlasAbstract{
A search for long-lived charginos produced either directly or in the cascade decay of heavy prompt gluino states is presented.
The search is based on proton--proton collision data collected at a centre-of-mass energy of $\sqrt{s}$~=~13~\TeV\ between 2015 and 2018 with the ATLAS detector at the LHC,
corresponding to an integrated luminosity of 136~fb$^{-1}$.
Long-lived charginos are characterised by a distinct signature of a short and then disappearing track, and are reconstructed using at least four measurements in the ATLAS pixel detector,
with no subsequent measurements in the silicon-microstrip tracking volume nor any associated energy deposits in the calorimeter.
The final state is complemented by a large missing transverse-momentum requirement for triggering purposes and at least one high-transverse-momentum jet.
No excess above the expected backgrounds is observed.
Exclusion limits are set at 95\% confidence level on the masses of the chargino and gluino for different chargino lifetimes.
Chargino masses up to 660~(210)~\GeV\ are excluded in scenarios where the chargino is a pure wino (higgsino).
For charginos produced during the cascade decay of a heavy gluino, gluinos with masses below 2.1~\TeV\ are excluded for a chargino mass of 300~\GeV\ and a lifetime of 0.2~ns.
 
}
 
\PreprintIdNumber{CERN-EP-2021-209}
 
\author{The ATLAS Collaboration}
 
\AtlasRefCode{SUSY-2018-19}

\hypersetup{pdftitle={ATLAS document},pdfauthor={The ATLAS Collaboration}}
 
\begin{document}
 
\maketitle

% The next lines are included from the .//sections/introduction.tex input file
 
\section{Introduction}
 
Supersymmetry\,(SUSY)\,\cite{Golfand:1971iw,Volkov:1973ix,Wess:1974tw,Wess:1974jb,Ferrara:1974pu,Salam:1974ig} is a space-time symmetry that extends the Standard Model (SM),
predicting the existence of a partner for each SM particle.
This extension provides solutions to deficiencies in the SM, including a candidate for dark matter\,(DM) as the lightest supersymmetric particle~(LSP) and a resolution to the hierarchy problem.
Superpartners of the SM particles have the same quantum numbers as their partner particles but differ by one half unit of spin.
Supersymmetric partners of the electroweak gauge bosons and the Higgs bosons, collectively referred to as electroweakinos,
consist of the bino, winos, and higgsinos, which mix to form neutral and charged mass eigenstates called neutralinos and charginos respectively.
The winos are the superpartners of the SU(2) gauge fields, the bino is the superpartner of the U(1) gauge field,
and the higgsinos are the superpartners of the Higgs fields.
 
The mass difference between the lightest neutralino\,(\ninoone) and the lightest chargino\,(\chinoonepm), $\mdiffchi$,
is predicted to be of the order of 100~\MeV, due to an essentially SM radiative correction~\cite{Ibe:2012sx,Fukuda:2017jmk}, in scenarios
where the LSP is  wino-like and other SUSY particles are decoupled.
In particular, anomaly-mediated supersymmetry breaking (AMSB) models\,\cite{Giudice:1998xp,Randall:1998uk} give rise to such differences and naturally predict a pure-wino LSP.
The mass-splitting between the charged and neutral wino in such scenarios is suppressed at tree level by the approximate custodial symmetry;
it has been calculated at the two-loop level to be around 160~\MeV\,\cite{Ibe:2012sx}, corresponding to charginos with $c\tau\sim$~58~mm ($\tau_{\chinoonepm}$~=~0.2~ns).
 
In addition to the wino LSP scenarios, a number of `natural' models of SUSY~\cite{Papucci:2011wy,Barbieri:1987fn,deCarlos:1993yy} predict a light higgsino LSP with a mass as light as the electroweak scale.
In these scenarios, the higgsino mass parameter $|\mu|$ is small compared to the other electroweak gaugino mass scales.
At tree level, the charged and neutral states are all mass degenerate, but
due to higher-order SM loop corrections a mass-splitting of approximately 300~\MeV{} is generated.
For chargino masses ranging from 91~\GeV{} to 1000~\GeV, the mass-splitting ranges from approximately 280~\MeV{} to 350~\MeV{}~\cite{PhysRevLett.81.34},
leading to charginos with $c\tau\sim$~14~mm ($\tau_{\chinoonepm}$~=~0.048~ns) to 7~mm ($\tau_{\chinoonepm}$~=~0.026~ns)~\cite{Fukuda:2017jmk}.

In both the wino and higgsino scenarios, the chargino can be produced with large momentum and live long enough to traverse multiple layers of the ATLAS pixel detector before decaying.
It decays primarily into a neutral weakly interacting LSP and a low-momentum pion.
Before it decays, the chargino deposits energy in the innermost tracking layers and can be reconstructed as a short track if at least four pixel layers are hit.
The weakly interacting LSP will escape detection and lead to missing transverse momentum,
while the pion from the chargino decay has insufficient momentum to be reconstructed as a track,
resulting in a characteristic signature where the track from the chargino disappears a short distance into the detector.
The algorithms used to reconstruct these disappearing tracks need at least four pixel hits
and therefore require charginos to have transverse path lengths that at least exceed the position of the fourth pixel layer at  $\sim$122~mm.
This makes the higgsino-like scenarios considerably more challenging than the wino-like models from an experimental perspective because of the extremely short chargino lifetime predicted in the higgsino models.
 
\begin{figure}[h]
\begin{center}
\subfloat[$pp\rightarrow \chinoonepm\ninoone j$]{\includegraphics[height=0.25\textheight]{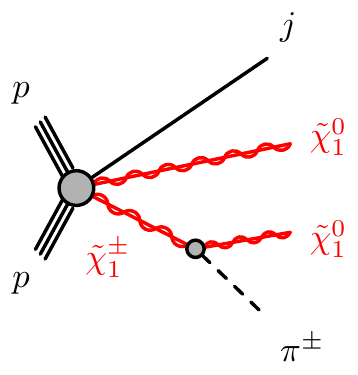}}
\subfloat[$pp\rightarrow qqqq\chinoonepm\ninoone$]{\includegraphics[height=0.23\textheight]{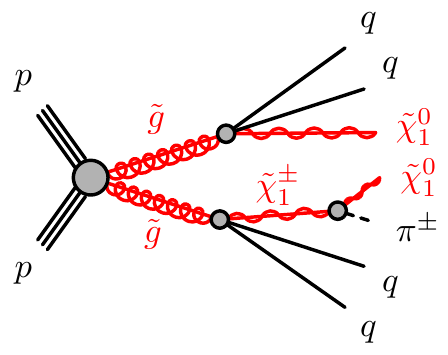}}
\caption{ Example diagrams for the (a) electroweak and (b) strong production channel signal models.
The signal signature consists of a long-lived chargino, missing transverse momentum, and quarks or gluons, which are observed as jets, and which
originate from (a) initial-state radiation or (b) in the cascade decay of the gluino.}
\label{fig:SUSYFeynman}
\end{center}
\end{figure}

This paper targets two production processes, the electroweak production of charginos and neutralinos,
and the strong production of gluinos after which charginos are produced during the cascade decay of the gluino, as shown in Figure~\ref{fig:SUSYFeynman}.
In both scenarios, the chargino is long-lived and reconstructed from energy deposits in the ATLAS pixel detector.
For the electroweak production process, a high-momentum jet from initial-state radiation~(ISR) is required in order to ensure that significant missing transverse momentum is available for event triggering.
The events selected from the electroweak and strong production channels are characterised by at least one and at least four jets, respectively, large missing transverse momentum, and at least one disappearing track with large transverse momentum.
 
In addition to the SUSY-specific wino and higgsino models studied in this paper,
the disappearing-track signature is typical of a large class of DM models that predict a DM thermal relic from a massive particle with only electroweak gauge interactions.
The wino and higgsino models probed in this search are part of a more generic class of models containing spin-1/2 particles transforming under SU(2) symmetry,
which give rise to a doublet or triplet of new particles.
The neutral mass eigenstate is a DM candidate, while the charged eigenstates give rise to the disappearing-track signature.
These DM models have gained interest in the wider community as an important signature to be considered at future colliders~\cite{Ellis:2691414}.
 
Previous searches for long-lived charginos resulting in a disappearing-track signature were performed by ATLAS~\cite{SUSY-2016-06,ATL-PHYS-PUB-2017-019}
using 36.1~\ifb{} of proton--proton collision data at a centre-of-mass energy of $\sqrt{s}=13$~\TeV{}.
The previous ATLAS results benefited from the addition of a new innermost pixel tracking layer which was installed at a radius of approximately 33~mm during the LHC long shutdown between Run~1 and Run~2.
The extra layer of pixels allowed the previous Run-2 analyses to reconstruct tracks shorter than those in the Run-1 analysis~\cite{SUSY-2013-01} and to be more sensitive to shorter chargino lifetimes.
The previous ATLAS results excluded winos with lifetimes of 0.2~ns for chargino masses up to 460~\GeV, and pure higgsinos for chargino masses up to 152~\GeV.
For gluino production, gluino masses up to 1.64~\TeV{} were excluded for an assumed chargino mass of 460~\GeV{} and 0.2~ns lifetime.
The CMS Collaboration has searched for long-lived charginos at a centre-of-mass energy of $\sqrt{s}=13$~\TeV.
For wino-like models, charginos were excluded with masses below 474~\GeV{} and a lifetime of 0.2~ns~\cite{CMS-EXO-19-010},
while for models with prompt gluinos that produce long-lived charginos in the decay of the gluino, CMS excluded gluinos up to 2.4~TeV~\cite{CMS:2019ybf}.
 
In this paper, the sensitivity to charginos with theoretically preferred wino and higgsino lifetimes is significantly improved due
to the much larger dataset corresponding to an integrated luminosity of 136~\ifb{} and new analysis methods.
The new analysis methods include updated signal region selection criteria and improved track-quality requirements.
The new track-quality criteria require the disappearing track to be isolated from calorimeter energy deposits.
This significantly enhances the rejection of dominant backgrounds
and accounts for two-thirds of the total improvement in the mass reach of the search in the electroweak pure-wino production channel.
 
The paper is structured as follows. A brief overview of the ATLAS detector is given in Section~\ref{sec:detector}.
Section~\ref{sec:dataset} provides details about the data samples, trigger, and simulated signal processes used in this analysis.
The reconstruction algorithms and event selection are presented in Sections~\ref{sec:objects} and \ref{sec:event_selection}, respectively.
Backgrounds are estimated in a fully data-driven manner and described in Section~\ref{sec:background_estimation}.
The systematic uncertainties are described in Section~\ref{sec:syst}.
The observed events in the signal and validation regions, and the statistical interpretation of the results are presented in Section \ref{sec:results}.
Section~\ref{sec:conclusion} is devoted to the conclusions.
% End of text imported from the .//sections/introduction.tex input file
% The next lines are included from the .//sections/detector.tex input file
\section{ATLAS detector\label{sec:detector}}
 
ATLAS\,\cite{PERF-2007-01} is a multipurpose detector with a forward--backward symmetric cylindrical geometry, covering nearly the entire solid angle around an interaction point of the LHC.\footnote{
ATLAS uses a right-handed coordinate system with its origin at the nominal interaction point in the centre of the detector.
The positive $x$-axis is defined by the direction from the interaction point to the centre of the LHC ring, with the positive $y$-axis pointing upwards, while the beam direction defines the $z$-axis.
Cylindrical coordinates ($r$, $\phi$) are used in the transverse plane, $\phi$ being the azimuthal angle around the $z$-axis.
The pseudorapidity $\eta$ is defined in terms of the polar angle $\theta$ by $\eta = - \ln\tan(\theta/2)$.
}
The inner tracking detector (ID) consists of pixel and microstrip silicon detectors covering the pseudorapidity region of $|\eta|<2.5$,
surrounded by a transition radiation tracker (TRT), which improves the momentum measurement and enhances electron identification capabilities.
The pixel detector spans the radius range from 3~cm to 12~cm, the microstrip semiconductor tracker (SCT) spans 30~cm to 52~cm, and the TRT spans 56~cm to 108~cm.
The pixel detector has four barrel layers and three disks in each of the forward and backward regions.
The barrel layers surround the beam pipe at radii of 33.3, 50.5, 88.5, and 122.5~mm, covering $|\eta|<1.9$.
These layers are equipped with pixels which have widths of 50~$\mu$m in the transverse direction.
The pixel sizes in the longitudinal direction are 250 $\mu$m for the first layer and 400~$\mu$m for the other layers.
The innermost layer, the insertable B-layer\,\cite{ATLAS-TDR-2010-19}, improves the reconstruction of tracks by adding an additional measurement close to the interaction point.
The ID is surrounded by a thin superconducting solenoid providing an axial 2~T magnetic field and by a fine-granularity lead/liquid-argon (LAr) electromagnetic calorimeter covering $|\eta| < 3.2$.
The calorimeters in the region of $3.1<|\eta|<4.9$ are made of LAr active layers with either copper or tungsten as the absorber material.
A steel/scintillator-tile calorimeter provides coverage for hadronic showers in the central pseudorapidity range of $|\eta|<1.7$.
LAr hadronic endcap calorimeters, which use copper as an absorber, cover the forward region of $1.5<|\eta|<3.2$.
The muon spectrometer~(MS) with an air-core toroid magnet system surrounds the calorimeters.
The ATLAS trigger system\,\cite{TRIG-2016-01} consists of a hardware-based level-1 trigger followed by a software-based high-level trigger.
An extensive software suite~\cite{ATL-SOFT-PUB-2021-001} is used in the reconstruction and analysis of real and simulated data, in detector operations, and in the trigger and data acquisition systems of the experiment.
% End of text imported from the .//sections/detector.tex input file
% The next lines are included from the .//sections/dataset.tex input file
\section{Dataset and simulated event samples\label{sec:dataset}}
 
The dataset for this search was collected at a centre-of-mass energy of $\sqrt{s}=13$~\TeV{} during Run~2 of the LHC between 2015 and 2018 by the ATLAS experiment.
The LHC collided protons at bunch-crossing intervals of 25~ns, with the average number of interactions per bunch crossing, $\mu$, ranging between 30 and 70 during the data-taking period.
 
Events are required to have been taken during stable beam conditions and when all the detector subsystems were operational.
In addition, luminosity blocks, defined to be periods of stable data-taking conditions typically 60 seconds long,
that have been identified to contain inactive SCT elements due to a power-supply crate trip, module desynchronisation, or readout problems are vetoed.
This data quality requirement is intended to protect against fake disappearing tracks arising from readout problems in the SCT detector, and results in a 2.7~\ifb{} loss of data.
 
After the application of additional requirements for beam and detector conditions, an inactive-SCT-element veto, and general data-quality requirements~\cite{DAPR-2018-01}, the total integrated luminosity is 136~\ifb.
The uncertainty in the combined 2015--2018 integrated luminosity is 1.7 \%~\cite{ATLAS-CONF-2019-021}, obtained using the LUCID-2 detector~\cite{LUCID2} for the primary luminosity measurements.

Events were collected using either missing transverse momentum (\met{}) or single-lepton triggers.
The selection thresholds of the \met{} triggers varied between 70~\GeV{} and 110~\GeV{} depending on the data-taking period~\cite{TRIG-2019-01},
while the single-lepton triggers required a single electron or muon with varying thresholds for the transverse momentum and isolation of the lepton~\cite{TRIG-2018-01,TRIG-2018-05}.
The \met{} triggers are the main triggers for the signal, control, and validation regions, while the single-lepton triggers are used for measurements of the \met{} trigger efficiency,
transfer factors for the background estimations, and the smearing functions.
 
While all backgrounds are determined in a data-driven manner,
samples of simulated Monte Carlo~(MC) events are used to cross check the background estimation and derive MC correction factors.
The production of single electroweak bosons plus jets was simulated with the
\SHERPA[2.2.1]~\cite{Bothmann:2019yzt}
generator using next-to-leading-order (NLO) matrix elements (ME) for up to two partons, and leading-order (LO) matrix elements
for up to four partons calculated with the Comix~\cite{Gleisberg:2008fv}
and \OPENLOOPS~\cite{Buccioni:2019sur,Cascioli:2011va,Denner:2016kdg} libraries. They
were matched with the \SHERPA parton shower~\cite{Schumann:2007mg} using the \MEPSatNLO
prescription~\cite{Hoeche:2011fd,Hoeche:2012yf,Catani:2001cc,Hoeche:2009rj}
and a set of tuned parameters developed by the \SHERPA authors.
Samples of diboson final states were simulated with the
\SHERPA[2.2.1] or 2.2.2~\cite{Bothmann:2019yzt} generator depending on the process,
including off-shell effects and Higgs boson contributions, where appropriate.
Fully leptonic final states and semileptonic final states, where one boson
decays leptonically and the other hadronically, were generated using
matrix elements at NLO accuracy in QCD for up to one additional parton
and at LO accuracy for up to three additional parton emissions.
The \NNPDF[3.0nnlo] set~\cite{Ball:2014uwa} of parton distribution functions (PDFs) was used for both single and diboson samples, along with the
dedicated set of tuned parton-shower parameters developed by the \SHERPA authors.
The production of \ttbar events was modelled using the
\POWHEGBOX[v2]~\cite{Frixione:2007nw,Nason:2004rx,Frixione:2007vw,Alioli:2010xd}
generator at NLO with the \NNPDF[3.0nlo]~\cite{Ball:2014uwa} PDF set
and the \hdamp parameter\footnote{The
\hdamp parameter is a resummation damping factor and one of the
parameters that controls the matching of \POWHEG matrix elements to
the parton shower and thus effectively regulates the
high-\pt radiation against which the \ttbar system recoils.} set
to 1.5\,\mtop~\cite{ATL-PHYS-PUB-2016-020}.  The events were interfaced
to \PYTHIA[8.230]~\cite{Sjostrand:2014zea} to model the parton shower,
hadronisation, and underlying event, with parameters set according
to the A14 tune~\cite{ATL-PHYS-PUB-2014-021} and using the \NNPDF[2.3lo]
set of PDFs~\cite{Ball:2012cx}. The decays of bottom and charm hadrons
were performed by \EVTGEN[1.6.0]~\cite{Lange:2001uf}.
Finally, single electron, muon and pion events are generated to study the simulated detector response.
The particles are generated uniformly in the range $0.1<\eta<1.9$ and with transverse momentum in the range of $10<p_{T}<100$~GeV.
 
Samples of simulated MC signal events were used to estimate the experimental sensitivity to various SUSY models.
The SUSY mass spectrum, branching ratios, and decay widths were calculated using {\scshape ISASUGRA}~v7.80~\cite{Paige:2003mg}.
The wino-like models were generated in the minimal AMSB model~\cite{Giudice:1998xp,Randall:1998uk} with $\tan \beta = 5$ and with a positive sign of the higgsino mass parameter.
The signal MC samples were generated using {\scshape MadGraph5}~v2.6.2~\cite{Alwall:2014hca} with up to two additional partons at leading order in the matrix element,
interfaced to \PYTHIA[8.230]~\cite{Sjostrand:2007gs} and \EVTGEN[1.6.0]~\cite{Lange:2001uf} for parton showering and hadronisation.
The CKKW-L merging scheme~\cite{Lonnblad:2011xx} was applied to combine the matrix elements with the parton shower.
The A14 tune~\cite{ATL-PHYS-PUB-2014-021} of \PYTHIA[8] was used with the \NNPDF[2.3lo] PDF set.
The response of the ATLAS detector to each MC event was modelled by a simulation based on \GEANT~\cite{Agostinelli:2002hh,SOFT-2010-01}.
For electroweak production, two (three) modes with at least one chargino are considered for wino (higgsino) models: $\chinoonepm\chinoonemp$, $\ninoone\chinoonepm$ (and $\chinoonepm\ninotwo$).
For the chargino decay in the higgsino model, branching ratios of 95.5\% for $\chinoonepm \rightarrow \pi^{\pm} \ninoone$, 3\% for $\chinoonepm \rightarrow e^{\pm}\nu \ninoone$
and 1.5\% for $\chinoonepm \rightarrow \mu^{\pm}\nu \ninoone$ are used~\cite{PhysRevLett.81.34},
while the AMSB model used a 100\% branching ratio for $\chinoonepm \rightarrow \pi^{\pm} \ninoone$.
For strong production, a simplified model is used: it assumes the branching ratios of the gluino decay are 1/3 for each of $\tilde{g} \rightarrow qq\tilde{\chi}^0_{1}$,
$\tilde{g} \rightarrow qq\tilde{\chi}^{-}_{1}$ and $\tilde{g} \rightarrow qq\tilde{\chi}^{+}_{1}$.
Only four flavours of quarks are considered for the strong production model: $d$, $u$, $c$ and $s$.
The proper lifetime of the chargino is an assumed free parameter that is scanned up to $\tau_{\chinoonepm}$ $<$10~ns.

The production cross-sections for the AMSB wino-like models are computed using {\scshape Prospino2}~\cite{Beenakker:1996ch} at NLO accuracy in the strong coupling constant.\footnote{A comparison with {\scshape Resummino} at NLO+NLL accuracy was performed and the deviations from the {\scshape Prospino2} result were found to be smaller than the combined scale and PDF uncertainties.}
The higgsino and strong production cross-sections are computed with {\scshape Resummino}~\cite{Debove:2010kf,PhysRevD.102.095021} at NLO plus next-to-leading-log (NLL) precision in the strong coupling,
and assuming mass-degenerate \ninoOneTwo{} and \chinoonepm{} higgsino sparticles.  All the other sparticles are assumed to be heavy and decoupled~\cite{Fuks:2012qx,Fuks:2013vua}.
The cross-sections for the higgsino models are approximately a factor of four smaller than in the AMSB wino-like models for electroweak production.
The nominal cross-section and its uncertainty are derived using the \PDFforLHC{} PDF set, following the recommendations of Ref. \cite{Butterworth:2015oua},
considering only first- and second-generation squarks $(\tilde{u}, \tilde{d}, \tilde{s}, \tilde{c})$.
Uncertainties due to the choice of renormalisation and factorisation scales are included by varying the scales from their nominal values by a factor of two or one half.
 
Inelastic $pp$ interactions modelled by \PYTHIA[8.186] and \EVTGEN[1.6.0] with the \NNPDF[2.3lo] PDF set
were overlaid onto the hard-scattering process to simulate the effect of pile-up.
The MC events were reweighted to match the distribution of the average number of interactions per bunch crossing observed in the data.
% End of text imported from the .//sections/dataset.tex input file
% The next lines are included from the .//sections/objects.tex input file
 
\section{Object reconstruction\label{sec:objects}}
 
Primary vertices are reconstructed from at least two tracks with transverse momentum~$\pt>500~\MeV$.
All vertices within the beam spot area are considered to be primary vertices.
The vertex with the largest summed $\pT^2$ of associated tracks is defined to be the hard-scattering vertex.
Events are required to have at least one hard-scatter vertex to be considered for analysis.

Electron candidates, used along with muons in the estimation of backgrounds and in the signal region event selection,
are reconstructed using energy clusters in the electromagnetic calorimeter which are matched to an inner detector track~\cite{EGAM-2018-01}.
They are required to have $\pt>10~\GeV${} and $|\eta|<2.47$ and must satisfy the `LooseAndBLayerLLH' quality criteria~\cite{PERF-2017-01}.
In order to ensure that the trajectories of the electrons are consistent with originating from the hard-scatter vertex,
the longitudinal impact parameter measured relative to the hard-scatter vertex ($z_{0}^{\text{HS}}$) must satisfy $|z_{0}^{\text{HS}}\sin\theta|<0.5$~mm,
and for the transverse impact parameter $d_{0}$, the significance $|d_{0}/\sigma_{d_{0}}|$ must be less than 5.
Electrons are required to be isolated from other objects by using a combination of track- and calorimeter-based information.
The sum of the transverse energy within a cone of size $\Delta R = \sqrt{ (\Delta\eta)^{2} + (\Delta \phi)^{2} } = 0.4$ around the electron candidate, divided by the electron's transverse energy,
is required to be less than 0.15 (0.20) for track-based (calorimeter-based) isolation criteria.
 
Muons are reconstructed by combining an inner detector track with a muon spectrometer track~\cite{MUON-2018-03}.
They are required to satisfy the `medium' quality requirements described in Ref.~\cite{MUON-2018-03} and have $\pt>10~\GeV${} and $|\eta|<2.7$.
Muon tracks that originate from the hard-scatter vertex of the event are selected by requiring
$|z_{0}^{\text{HS}}\sin\theta|<0.5$~mm and $|d_{0}/\sigma_{d_{0}}|<3$.
The muons must satisfy the same isolation requirements as applied to electrons.
Muons reconstructed by using only information from the muon spectrometer are used to estimate muon backgrounds as described in Section~\ref{sec:background_estimation:lepton}, and are referred to as stand-alone muons.
No constraint is placed on track impact parameters during the reconstruction of stand-alone muons.
 
Jet candidates are reconstructed from three-dimensional topological energy clusters~\cite{JETM-2018-05}
by using the anti-$k_{t}$ algorithm~\cite{Cacciari:2008gp,Fastjet} with radius parameter $R=0.4$.
Jets are corrected to particle level by application of a jet energy scale (JES) calibration derived from simulation
and by in situ corrections obtained from 13~\TeV{} data~\cite{JETM-2018-05}.
Jets are required to have $\pt>20~\GeV${} and $|\eta|<2.8$.
In order to reduce contributions from pile-up jets, all jets with $|\eta|<2.5$ and $\pt{}<60$~\GeV{} are required to satisfy the jet-to-vertex tagger (JVT)~\cite{PERF-2014-03} requirements.
The tagger is configured to have a 92\% efficiency to identify jets from the hard-scatter vertex.
 
An overlap removal procedure is applied to all objects to avoid double counting.
If an electron and a jet are separated by $\Delta R<0.2$, the electron candidate is kept and the jet is discarded.
For jets surviving this requirement, if an electron or muon is separated from the jet by $\Delta R < 0.4$,
the jet is kept and the electron or muon is discarded.
 
The missing transverse momentum is reconstructed as the negative vector sum of the transverse momenta of photons, electrons, muons and jets, and a soft term.
The soft term is reconstructed from tracks that are associated with the hard-scatter vertex
but not with any object already counted~\cite{PERF-2016-07,ATLAS-CONF-2018-023}.
 
The primary ATLAS tracking algorithm reconstructs each track from a combination of at least seven hits in the pixel and SCT detectors~\cite{ATL-PHYS-PUB-2009-002,ATL-PHYS-PUB-2015-018}.
The majority of simulated charginos decay before passing through enough detector layers to satisfy the minimum number of silicon detector hits required for track reconstruction.
To find these chargino tracks, a second-pass track reconstruction is performed after masking the hits used by the primary track reconstruction to avoid the loss of efficiency that would result from mis-reconstructed tracks that accidentally include hits from other charged particles.
The second-pass track reconstruction is seeded by four hits in the innermost pixel system, and the tracks are extended into the SCT and TRT detectors.
Such tracks are referred to as \emph{pixel tracklets} and are reconstructed from at least four hits in the ATLAS pixel detector.
 
Pixel tracklets are required to satisfy a series of quality and selection criteria in order to be distinguished from fake tracklets.
The selection of pixel tracklets was developed specifically for this search and is optimised to maintain high signal selection efficiency and good background rejection.
In addition to the requirement of four pixel hits on different pixel layers, the pixel tracklet must have a pixel hit on the innermost pixel layer and
must not have any hits that deviate significantly from the tracklet's trajectory. All four pixel hits must be recorded on consecutive pixel layers.
 
The chi-squared probability for each tracklet, which is calculated from fit-result locations and the measured hit locations, must be greater than 0.1.
The transverse momentum and the pseudorapidity of the tracklet must satisfy $\pt>20~\GeV${} and $0.1<|\eta|<1.9$, respectively.
Impact parameter requirements are applied to the pixel tracklets to ensure that they originate from the hard-scatter interaction.
The requirements are $|d_{0}/\sigma_{d_{0}}|<1.5$ and  $|z_{0}^{\text{HS}}\sin\theta|<0.5$~mm.

Finally, the tracklet is required to be isolated from other tracks in the event:
the sum of track momenta within $\Delta R=0.4$ of the pixel tracklet is required to be less than 4\% of the tracklet momentum.
Overlap removal is applied to the tracklet: if an electron or a stand-alone muon track is separated from it by $\Delta R<0.4$, the tracklet candidate is discarded.
 
The pixel tracklets for the signal events are characterised by an absence of hits in the outermost silicon layers and no associated calorimeter activity.
Together, these two requirements define the \emph{disappearing-track} conditions.
The first condition is enforced by a veto on pixel tracklets with any SCT hits.
The second condition is a new selection criterion developed after the previous ATLAS analyses~\cite{SUSY-2016-06,ATL-PHYS-PUB-2017-019}.
Since electron and hadron backgrounds tend to deposit a significant amount of energy in the calorimeters,
limiting the calorimeter energy allowed along the trajectory of the pixel tracklet is effective in separating signal from the electron and hadron backgrounds.
 
Due to the short lever arm of pixel tracklets, their momentum and spatial resolutions are worse than those of tracks from the primary track reconstruction.
For charginos with $\pt > 60~\gev$, the $\qoverpt$ resolution, where $q$ is the electric charge of the candidate chargino, and the azimuthal angle resolution of pixel tracklets are measured in simulation to be
$8.28 \pm0.05$~$\tev^{-1}$ and $0.4065\pm0.0027$~mrad, respectively.
Similarly, chargino tracks reconstructed in the primary reconstruction and having at least seven SCT hits are measured to have $\qoverpt$ and azimuthal angle resolutions of
$0.532\pm0.005$~$\tev^{-1}$ and $0.0801\pm0.0008$~mrad, respectively.
 
The calorimeter energy $E_{\textrm{T}}^{\textrm{topo cluster}}$ is calculated from the sum of topological energy clusters in the calorimeter whose angular separation from the pixel tracklet is $\Delta R < 0.2$.
The pointing resolution of the pixel tracklets in both $\eta$ and $\phi$ is found to be better than 1~mrad in all regions of the detector.
A requirement of $E_{\textrm{T}}^{\textrm{topo cluster}} < 5~\GeV{}$ is applied to all pixel tracklets.
Figure~\ref{fig:event_selection:calo_veto} shows a comparison of calorimeter energies for the simulated signal and the data-driven background predictions obtained as described in Section~\ref{sec:background_estimation}.
 
\begin{figure}[t]
\begin{center}
\includegraphics[width=0.80\textwidth]{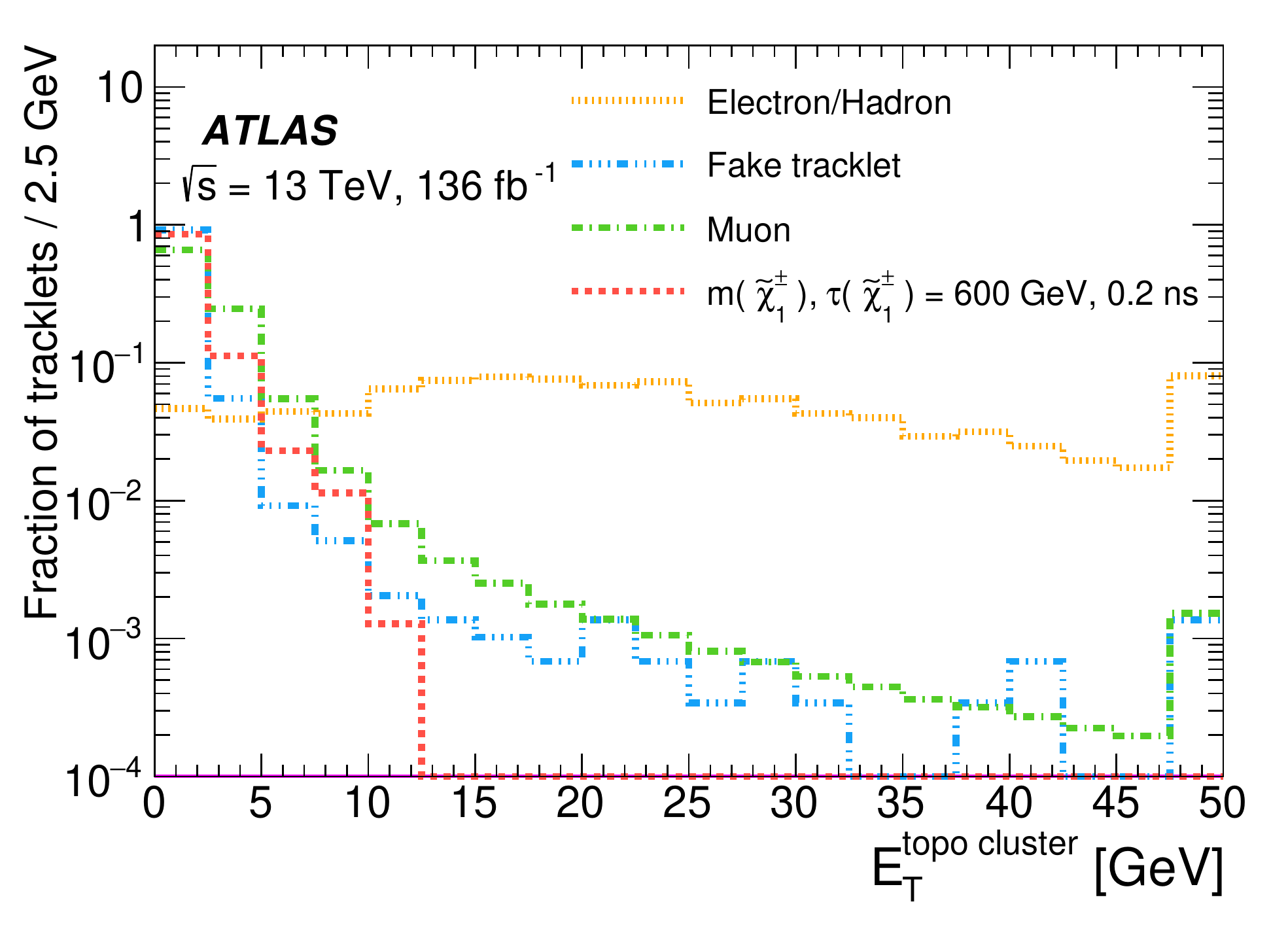}
\caption{Distribution of calorimeter energy found within $\Delta R=0.2$ of the pixel tracklet for a simulated signal
with $m_{\chinoonepm}= 600$~\GeV{} and $\tau_{\chinoonepm} = 0.2$~ns.
Backgrounds arising from scattering electrons, hadrons and muons, and combinatorial fake tracklets, are also shown.
Background predictions are obtained from data; the electron and hadron components are derived
from events with tracklet $\pt < 60$~\GeV{} and $\met > 100~\GeV$ with no requirement on $E_{\mathrm{T}}^{\textrm{topo cluster}}$.
All predictions are normalised to unity.
The last bin includes entries from the overflow bins.
}
\label{fig:event_selection:calo_veto}
\end{center}
\end{figure}
 
% End of text imported from the .//sections/objects.tex input file
% The next lines are included from the .//sections/event_selection.tex input file
\section{Signal region selection\label{sec:event_selection} }

Signal events for this analysis are characterised by a distinctive signature composed of a disappearing track and large missing transverse momentum.
The disappearing-track criteria are implemented by vetoing pixel tracklets with any SCT hits or significant calorimeter energy along their trajectory as described in the previous section.
Signal regions are designed to target the electroweak and strong production models shown separately in Figure~\ref{fig:SUSYFeynman}.
 
A common preselection is applied to all events selected for the signal regions.
Events are required to contain at least one high-momentum jet with $\pt>100~\gev${} and satisfy the \met{} trigger requirements described in Section~\ref{sec:dataset}. 
In events containing multiple pixel tracklets satisfying all of the quality criteria described in Section~\ref{sec:objects}, the tracklet with the highest transverse momentum is chosen.
In order to reduce contributions from backgrounds such as top-pair and $W/Z+\mathrm{jets}$ production,
events are vetoed if they contain any electron or muon candidates.
 
In this analysis, the \met{} trigger efficiency is measured in data and used instead of the simulated trigger's efficiency.
In order to evaluate the \met{} trigger efficiency for signal-like events, a control sample consisting of data events with exactly one muon, with $\pt>27~\gev$, matched to the single-muon trigger is used.
Muon candidates are not included as visible objects in the hardware-based level-1 \met{} trigger, so $W \to \mu\nu$ events, which have a signature similar to that of signal events, can be used to assess the trigger performance.
In order to select events consistent with a $W \to \mu\nu$ decay, events are required to have  $30~\GeV<m_{\mathrm T}<100~\GeV$, where $m_{\mathrm T}$ is the transverse mass defined by
 
\begin{equation}
m_{\mathrm T}=\sqrt{2p_{\mathrm T}^{\mathrm \mu} \met \left[ 1-\cos \Delta \phi ({\mu}, \met)\right] }. \nonumber
\end{equation}
 
The \met{} trigger efficiency is calculated by dividing the number of these events passing the \met{} trigger by the total number of these events.
The efficiencies depend on the trigger thresholds used during a given data-taking period, and all plateau at \met{} values above 200~\gev.
In the signal regions where the \met{} requirements are greater than 200~\gev{} or 250~\gev{}, as described below, the trigger efficiency is nearly 100\%.
The efficiencies are determined in bins of \met with widths of 10~\gev.
 
In order to factor out efficiency differences between data and simulation,
all simulated-signal predictions use the efficiencies measured in data.
For selections with \met{} below 200~\gev, such as those for the control and validation regions described in Section~\ref{sec:evtSel_CRs},
the event selection efficiency is significantly impacted by the trigger efficiency turn-on.
The efficiency differences between data and simulation do not affect the background estimations because the efficiencies are based on data satisfying the same \met{} trigger requirements, and the contamination of the control regions by signal events is expected to be small.
 
Signal regions are designed for the electroweak and strong production modes.
For the signal regions targeting the electroweak (strong) production mode,
the offline missing transverse momentum is required to be greater than 200~(250)~\GeV.
The two signal regions are not orthogonal.
The higher \met{} requirement is found to be optimal for the strong production channel, due to the large Lorentz boost of the neutralinos produced in the decay of a heavy gluino.
Additionally, the strong production signal region requires at least three jets with $\pt>20~\gev$.
In order to reduce contributions from processes for which fake instrumental \met{} can be produced by jet mismeasurements, the azimuthal angle $\Delta \phi^{{\text{jet}}-\met}_{\mathrm{min}}$,
defined as the smallest of the $\phi$ direction differences between the missing transverse momentum
and each of the four highest-\pt{} jets with $\pt>50~\gev$, is required to be greater than 1.0 (0.4) for the electroweak (strong) production signal regions.
The signal region selection criteria are summarised in Table~\ref{tab:SR:selectionSummary}.
 
The transverse momentum distribution of the selected pixel tracklets is fitted to estimate the backgrounds and sensitivity to the signals.
Since the signal is characterised by a high-momentum chargino, the majority of the signal events contain a high-\pt{} pixel tracklet.
For the model-independent interpretation of the results, the pixel tracklet is required to have $\pt>60~\gev${} as shown in Section~\ref{sec:results}.
 
\begin{table}[thbp]
\small
\caption{Signal region selection for electroweak and strong production channels. Entries with `--' indicate that no requirement is placed on the variable.}
\label{tab:SR:selectionSummary}
\begin{center}
\begin{tabular}{lcc}
\toprule
& \hspace*{30mm} \\[-3mm]
{Signal region}                                           & {Electroweak production }        & {Strong production}     \\
\midrule
Number of electrons and muons                & \multicolumn{2}{c}{~~~~$0$}                                        \\
Number of pixel tracklets                                        & \multicolumn{2}{c}{$\geq 1$}                            \\
\midrule
\met[\gev]                     & $>200$                  & $>250$             \\
Number of jets~($\pt>20~\gev$)    & ~~~~$\geq 1$                 & ~~~~$\geq 3$            \\
Leading jet \pt{} [\gev]              & $> 100$                            & $>100$       \\
Second and third jet \pt{} [\gev]          & --                            & ~~$>20$  \\
$\Delta \phi^{{\text{jet}}-\met}_{\mathrm{min}}$  (up to 4$^{\text{th}}$ jet with $\pt{}>50$~GeV)         & $>1.0$                           & $>0.4$  \\
\bottomrule
\end{tabular}
\end{center}
\end{table}
 
Table~\ref{table:TrackCutFlow_Signal} shows the expected number of signal events, normalised to 136~\ifb{}, satisfying the selections of the electroweak and strong production channels.
The electroweak and strong production signals assume $m_{\chinoonepm}=600~\GeV{}$ and $(m_{\tilde{g}},m_{\chinoonepm})=(1.4,1.1)~\TeV{}$ respectively.
Unless otherwise indicated, the wino LSP model is assumed for the benchmark signal points.
The dominant inefficiency in the pixel tracklet selection arises from the requirement that the tracklet can be reconstructed,
and thus has propagated at least to the fourth pixel layer at a radius of 122.5~mm.
Such a requirement implies that charginos have a significant Lorentz boost.
The overall acceptance and efficiency for those benchmark signal points in the case of a chargino lifetime of 0.2~ns are
$0.74$\% (1.5\%) and $40$\% (37\%), respectively, for the electroweak (strong) production channel.
The acceptance is calculated by counting the number of events passing the event section in Table~\ref{tab:SR:selectionSummary} at generator level.
Additionally, at least one chargino is required to have a decay radius after the fourth pixel layer ($r>122.5$~mm) and before the first SCT layer or end-cap (r$<299$~mm, $|z|<934$~mm),
transverse momentum greater than 60~GeV, its pseudorapidity $0.1<|\eta|<1.9$, and an angular separation of $\Delta R>0.5$ for the four highest \pt{} jets with transverse momentum greater than 50~GeV.
The efficiency term accounts for any additional effects at reconstructed level, including the trigger and pixel tracklet reconstruction inefficiencies.
 
\begin{table}[thbp]
\small
\caption{
Expected number of signal events after each of the selection requirements listed in Section~\ref{sec:event_selection} for the electroweak and strong production channels as well as for two different chargino lifetimes.
Signal predictions are normalised to an integrated luminosity of 136~\ifb.
Errors are statistical only.}
\label{table:TrackCutFlow_Signal}
\centering
\begin{tabular}{lcccc}
\toprule
Signal production channel                                & \multicolumn{2}{c}{Electroweak production}  & \multicolumn{2}{c}{Strong production}  \\
$\tau_{\chinoonepm}$                                     & \multicolumn{1}{c}{0.2 ns} & \multicolumn{1}{c}{1.0 ns}  & \multicolumn{1}{c}{0.2 ns} & \multicolumn{1}{c}{1.0 ns} \\
\midrule
\met{} trigger                                           & 770.8 $\pm$  6.8 & 775.3 $\pm$ 5.2 & 3177 $\pm$ 22  & 3177 $\pm$ 22      \\
Lepton veto                                              & 769.4 $\pm$  6.8 & 774.2 $\pm$ 5.2 & 3165 $\pm$ 22  & 3165 $\pm$ 22      \\
$\met > 200$~$\GeV$                                      & 394.5 $\pm$  5.2 & 390.9 $\pm$ 4.0 & \multicolumn{1}{c}{--}          & \multicolumn{1}{c}{--}                \\
$\met > 250$~$\GeV$                                      & \multicolumn{1}{c}{--}        & \multicolumn{1}{c}{--}          & 1852 $\pm$ 17  & 1852 $\pm$ 17      \\
Leading jet $\pt > 100~\GeV$                                 & 389.7 $\pm$  5.2 & 384.9 $\pm$ 4.0 & 1848 $\pm$ 17  & 1848 $\pm$ 17      \\
Third jet $\pt >  20~\GeV$                                 & \multicolumn{1}{c}{--}        & \multicolumn{1}{c}{--}          & 1834 $\pm$ 17  & 1834 $\pm$ 17      \\
$\Delta\phi_{\mathrm{min}}^{\mathrm{jet}-\met}> 1.0$     & 366.7 $\pm$  5.0 & 362.3 $\pm$ 3.9 & \multicolumn{1}{c}{--}          & \multicolumn{1}{c}{--}   \\
$\Delta\phi_{\mathrm{min}}^{\mathrm{jet}-\met}> 0.4$     & \multicolumn{1}{c}{--}        & \multicolumn{1}{c}{--}          & 1578 $\pm$ 16  & 1578 $\pm$ 16      \\
\midrule
Pixel tracklet selection ($\pt > 60~\GeV$)                  &  ~~~~8.6 $\pm$  0.6  & ~~27.3 $\pm$ 0.8  & ~~16.0 $\pm$  1.3 & 105.0 $\pm$  3.3  \\
 
\bottomrule
\end{tabular}
\end{table}
 
% End of text imported from the .//sections/event_selection.tex input file
% The next lines are included from the .//sections/background_estimation.tex input file
 
\clearpage
\section{Background estimation\label{sec:background_estimation}}
 
Backgrounds arise from several sources
and can be classified into two categories: \textit{charged-particle scattering} and \textit{combinatorial fake} backgrounds.
All backgrounds are estimated in a fully data-driven manner using an unbinned likelihood fit of background templates to the pixel tracklet \pt{} spectrum as described in Section~\ref{sec:background_estimation:fitting}.
Backgrounds templates are derived from data for the charged-particle scattering and combinatorial fake backgrounds
as described in Section~\ref{sec:background_estimation:lepton} and~\ref{sec:background_estimation:fake}, respectively.
The likelihood fit is then performed simultaneously in background enriched control regions (CRs) and the signal region.
Additional fits are performed with the validation regions in place of the signal region for the purpose of validating the background modelling.
A schematic representation of the various background processes contributing to this analysis,
as well as the expected signal signature, is shown in Figure~\ref{fig:signal_background_image}.
 
The charged-particle scattering background arises primarily from events in which a lepton changes its direction after traversing the inner pixel layers,
through interaction with the material or bremsstrahlung, so that its reconstructed track does not have any associated hits in the SCT and TRT detectors.
The dominant underlying processes contributing to the scattering backgrounds are $W\rightarrow \ell \nu$ and $t\bar{t}$ production.
Contributions from electrons, muons and charged hadrons are estimated separately.
The latter includes all physics processes that generate a hadronic final state, such as jets originating from the hadronisation of quarks and gluons.
Transfer factors, defined in Section~\ref{sec:background_estimation:lepton}, are measured in $Z \rightarrow \ell\ell$ events and applied to the electron,
muon and hadron CRs in order to estimate the relative compositions of the templates for the charged-particle scattering backgrounds.
In the CR, a lepton or inner-detector track is used as a proxy for the pixel tracklet,
and its momentum is smeared to match the pixel tracklet's transverse momentum as described in Section~\ref{sec:background_estimation:smearing}.
 
Combinatorial fake backgrounds arise from random combinations of unassociated pixel hits that are in close proximity and reconstructed into a pixel tracklet.
The shape of the combinatorial fake background is estimated using the fake-enriched high-$|d_{0}|$ CRs described in Section~\ref{sec:evtSel_CRs}.
The relative contributions of the charged-particle scattering and combinational fake backgrounds are constrained in a combined fit as described in Section~\ref{sec:background_estimation:fitting}.
 
\begin{figure}[t]
\begin{center}
\includegraphics[width=0.88\textwidth]{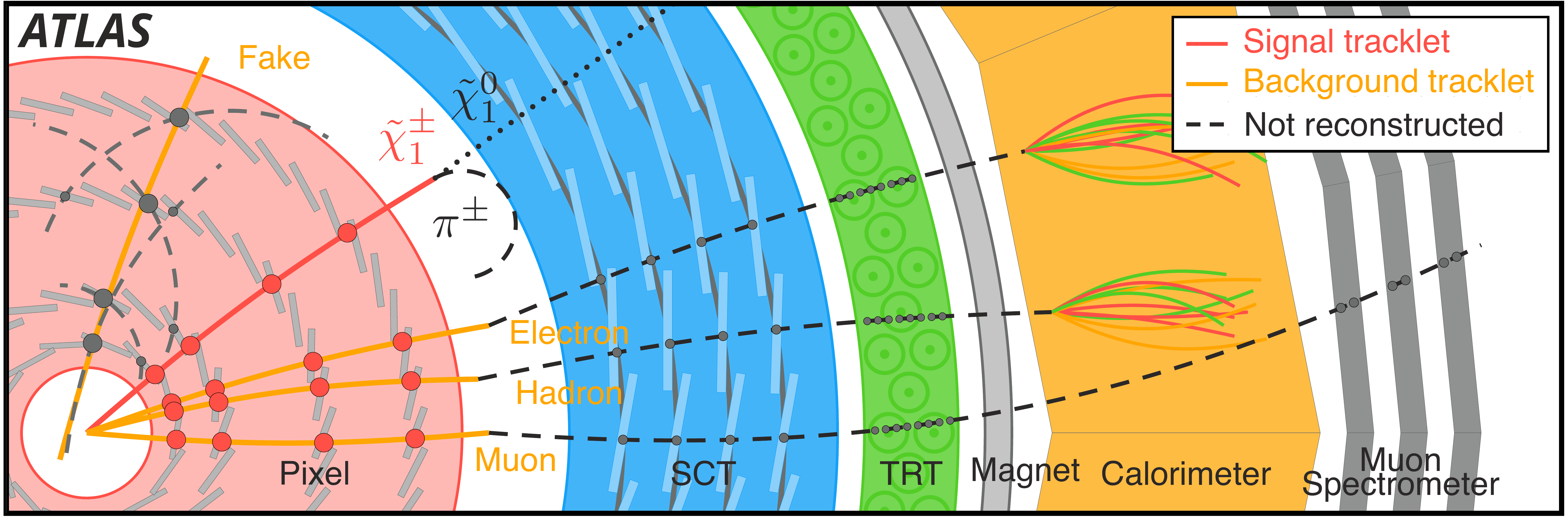}
\caption{ A pictorial representation of signal and background processes.
Detectors are not to scale and for illustration purposes only.
The signal chargino~($\chinoonepm$) decays into a charged pion~($\pi^{\pm}$) and neutralino~($\chinoonezero$).
`Fake', `Muon' and `Electron or hadron' in the figure represent the fake-tracklet background, the muon background and electron or hadron backgrounds respectively.
}
\label{fig:signal_background_image}
\end{center}
\end{figure}
 
\subsection{Control and validation regions\label{sec:evtSel_CRs}}
 
In order to estimate scattering backgrounds, low-\met{}, middle-\met{}, and high-\met{} CRs are defined.
Events selected for the CRs must satisfy the \met{} trigger requirements and pass all kinematic selections of the signal regions except the pixel tracklet requirement.
The pixel tracklet selection is replaced with a well-identified electron, muon or inner-detector track,
in order to estimate contributions from electron, muon or hadron scattering backgrounds, respectively.
The leading electron or muon is excluded from the calculation of the missing transverse momentum,
in order to reproduce the treatment of the pixel tracklets when calculating the offline \met{}.
Energy deposited by electrons in the calorimeters may be included in the \met{} calculation via other objects in the event,
but is treated in a similar manner for events with pixel tracklets or when electrons are excluded from the offline \met{} calculation.
The low- and middle-\met{} requirements are $100~\GeV<\met{}<150~\GeV${} and $150~\GeV<\met<200~\GeV$, respectively.
The high-\met{} requirements are the same as for the signal region for which the backgrounds are being estimated:
$\met>200~\GeV${} and $\met>250~\GeV${} for the electroweak and strong production signal regions, respectively.
Events in the CRs are used to calculate the number of pixel tracklets from scattering processes as described in Section~\ref{sec:background_estimation:lepton}.
 
A high-purity control sample of fake pixel tracklets is obtained by applying the same kinematic selection requirements as in the signal regions,
but with an inverted requirement of $ |d_0 / \sigma_{d_0}| > 10 $ on the transverse impact parameter.
Events selected for the fake-tracklet CR must satisfy the \met{} trigger requirements, but to increase the sample size no offline $\met$ selection is applied.
Events in the fake-tracklet CR are used to estimate the tracklet \pt{} shape and number of fake pixel tracklets in the signal regions as described in Section~\ref{sec:background_estimation:fake}.
 
A low-\met{} pixel-tracklet CR is used in the simultaneous fit described in Section~\ref{sec:background_estimation:fitting}.
This CR has the same selection as the signal region, except for the \met{} requirement to enrich in background contributions.
Validation regions (VR) defined in the context of the middle-\met{} selection are used to validate the background model constructed for the fit.
A low-\pt{} pixel tracklet selection is used to validate the background predictions.
In the low-\pt{} VR, the pixel tracklet is required to have $\pt<60~\GeV${} in order to reduce the contamination from signal at high transverse momentum.
In order to validate the fit results across the entire \pt{} spectrum in a region with low signal contamination, a middle-\met{} calorimeter-based sideband validation region is used.
The calorimeter sideband selection requires the pixel tracklet to have $5~\GeV<E_{\mathrm{T}}^{\textrm{topo cluster}}<10~\GeV$.
The number of signal events in the control and validation regions is less than one quarter of the estimated background uncertainty.
 
A summary of the control and validation regions is shown in Table~\ref{tab:fitting:CRs}.
 
\begin{table}[t]
\caption{Definitions of control and validation regions.
The track condition corresponds to the modified selection on the pixel tracklet, or tracks used as a proxy for the pixel tracklet to estimate background processes.
The table shows the treatment of the \met{} for the estimation of the scattering backgrounds.
The electron, muon and hadron control regions are defined for each of the low-, middle- and high-\met{} selections as described in the text. }
\label{tab:fitting:CRs}
\begin{center}
\begin{tabular}{ccc}
\toprule
Region & \met{} condition & Track condition \\
\midrule
\multicolumn{3}{c}{\textbf{Control regions}} \\
\midrule
Electron                        & $e$-subtracted \met{}      & well-identified electron \\
Muon                            & $\mu$-subtracted \met{}    & well-identified muon \\
Hadron                          & standard \met{}            & inner-detector track \\
Fake-tracklet                   & standard \met{}            & pixel tracklet with $|d_0 / \sigma_{d_0}| > 10$ \\
\midrule
\multicolumn{3}{c}{\textbf{Low-\met{} control region}} \\
\midrule
Low-\met{}                      & standard \met{}            & pixel tracklet \\
\midrule
\multicolumn{3}{c}{\textbf{Middle-\met{} validation regions}} \\
\midrule
Low-\pt{}                       & standard \met{}            & pixel tracklet with $\pt<60~\GeV${} \\
Calorimeter sideband           & standard \met{}            & pixel tracklet with $5~\GeV<E_{\mathrm{T}}^{\textrm{topo cluster}}<10~\GeV${} \\
\bottomrule
\end{tabular}
\end{center}
\end{table}
 
\subsection{Smearing functions\label{sec:background_estimation:smearing}}
 
The transverse momentum resolution of a track scales as $\Delta \pt / \pt \propto 1/L_{\mathrm{T}}^{2}$, where $L_{\mathrm{T}}$ is the transverse length of the track.
Since pixel tracklets have only pixel hits, they are significantly shorter than tracks with a full set of pixel, SCT and TRT hits,
and thus the transverse momentum resolution of the pixel tracklets is significantly worse than that of full-length tracks.
Therefore, the tracks in the CRs need to be smeared to match the pixel tracklet momentum resolution.
Additionally, the tracklet $\qoverpt$ resolution that is measured using observed data samples is worse than that predicted by simulation,
and measurements in data are used to correct those from simulation.
 
The $\qoverpt$ resolution has a strong dependence on the momentum of the track: at low momentum, multiple scattering effects are dominant,
while at higher momentum the alignment of the detector can affect the resolution.
Since a fit is applied to the shape of the pixel tracklet \pt{} spectrum, it is important to have a good description of the different components of the spectrum.
The measured tracklet resolution in data is used to correct the shapes of the track \pt{} templates in the CRs,
and additionally to correct the expected $\qoverpt$ resolution of the simulated-signal predictions to match the data.
 
Smearing functions are derived using $Z\rightarrow \mu\mu$ and  $Z\rightarrow ee$ events.
Events are required to satisfy the data quality criteria described in Section~\ref{sec:dataset},
and contain two opposite-sign, same-flavour leptons with an invariant mass between 81~\GeV{} and 101~\GeV.
The transverse momenta of the leading and sub-leading leptons are required to be greater 25~\GeV{} to satisfy the single lepton trigger requirements.
A specialised track reconstruction configuration is then used to re-fit the leptons using pixel hits only,
and the re-fitted leptons are used to derive a $\qoverpt$ smearing function with respect to the lepton four-momentum.
 
The $\Delta(\qoverpt)$ distribution is fitted using the ExpGaussExp function~\cite{Das:2016stf}:
 
\begin{eqnarray}
f(z; \sigma,\alpha) &=&
\begin{cases}
\exp( \alpha (z + \alpha/2) ) & (z < -\alpha) \\
\exp( - z^2/2 )                 & (-\alpha < z < \alpha), \\
\exp(-\alpha (z - \alpha/2) ) & (z > \alpha) \\
\end{cases} \\
z &=& \frac{\Delta (\qoverpt) - \beta}{\sigma},
\label{eq:CrystalballLikeFunction}
\end{eqnarray}
 
\noindent
where $\alpha$ is a parameter that reflects the slope of the tail, and $\beta$ and $\sigma$ are parameters that control the mean and resolution, respectively, of the core of the distribution.
The parameter $\sigma$ has a significant \pt{} dependence, and parameters are measured in data for $\pt > 25~\GeV$ using data collected with single lepton triggers.
Below this threshold, the smearing functions are estimated using detector simulation for tracks with transverse momentum down to $\pt = 10~\GeV$.
The parameters derived above $\pt > 25~\GeV$ agree well between data and detector simulation up to a constant overall difference of approximately 10\% across all track momenta.
This difference is used to correct the parameters derived with the detector simulation for tracks with $10<\pt < 25~\GeV$ .
The uncertainties on the $\sigma$ and $\alpha$ terms are 8.2\% and 9.4\% for muons and 8.4\% and 11.5\% for electrons, respectively.
These uncertainties include statistical errors added together in quadrature with the maximum deviation of the smearing parameters across different data taking conditions.
The \pt{}-dependent smearing function parameters derived for electrons and muons are shown in Table~\ref{tab:smearing_parameters} along with their uncertainties.
These values are used for the full data set.
 
\begin{table}[thp]
\caption{The \pt{}-dependent $\sigma$ and $\alpha$ smearing parameters for muons and electrons.
The uncertainties include both statistical and systematic sources. }
\begin{center}
\begin{tabular}{ccccc}
\toprule
& \multicolumn{2}{c}{Muon} & \multicolumn{2}{c}{Electron} \\
\midrule
Transverse momentum [\gev] & $\sigma ~[\TeV^{-1}]$ & $\alpha$ & $\sigma ~[\TeV^{-1}]$ & $\alpha$   \\
\midrule
$10  < \pt < 15 $                     & $17.0 \pm1.4 $ & $1.7 \pm 0.1 $ & $20.9 \pm1.8 $ & $1.9 \pm 0.2 $ \\
$15  < \pt < 20 $                     & $15.5 \pm1.3 $ & $1.7 \pm 0.1 $ & $19.5 \pm1.6 $ & $1.9 \pm 0.2 $ \\
$20  < \pt < 25 $                     & $14.9 \pm1.2 $ & $1.7 \pm 0.1 $ & $18.3 \pm1.5 $ & $1.9 \pm 0.2 $ \\
$25  < \pt < 35 $                     & $14.8 \pm1.2 $ & $1.7 \pm 0.1 $ & $17.0 \pm1.4 $ & $1.9 \pm 0.2 $ \\
$35  < \pt < 45 $                     & $14.2 \pm1.2 $ & $1.7 \pm 0.1 $ & $15.4 \pm1.3 $ & $1.8 \pm 0.2 $ \\
$45  < \pt < 60 $                     & $13.6 \pm1.1 $ & $1.6 \pm 0.1 $ & $14.5 \pm1.2 $ & $1.7 \pm 0.2 $ \\
~~$60  < \pt < 100$                & $13.4 \pm1.1 $ & $1.7 \pm 0.1 $ & $13.9 \pm1.2 $ & $1.5 \pm 0.2 $ \\
$100 < \pt~~~~~~~~~~~$      & $13.2 \pm1.1 $ & $1.6 \pm 0.1 $ & $14.0 \pm1.2 $ & $1.6 \pm 0.2 $ \\
\bottomrule
\end{tabular}
\end{center}
\label{tab:smearing_parameters}
\end{table}
 
\subsection{Charged-particle background templates\label{sec:background_estimation:lepton}}
 
Backgrounds arising from charged particles are estimated using a fully data-driven method.
Transfer factors are derived using a $Z\rightarrow \ell\ell$ tag-and-probe method.
In this method, a good-quality tag electron or muon with $\pt>30~\GeV${} that is matched to any of the single-lepton triggers described in Section~\ref{sec:dataset} is selected for analysis.
Probe leptons are identified using clusters reconstructed from energy deposits in the calorimeters,
a muon track reconstructed only in the muon spectrometer, or a high-quality inner-detector track, depending on the measurement as described below.
Transfer factors are defined so as to represent the probability that a lepton is misidentified as a pixel tracklet and fulfils the full set of pixel tracklet selection criteria.
They are used to scale the number of events in the electron or muon CRs defined in Table~\ref{tab:fitting:CRs} to obtain an estimate of the number of events from charged-particle scattering in the signal regions.
 
All probe leptons are required to have $\pt>10~\GeV${} and $|\eta|<2.5$.
In order to select events originating from $Z\rightarrow \ell^{\pm}\ell^{\mp}$ decays, the probe lepton and tag lepton are required to have opposite-sign charges,
and the invariant mass of the tag-and-probe pair is required to satisfy $81~\GeV< m_{\text{tag,probe}} < 101~\GeV$.
For both the electron and muon transfer factors, the charge of the probe lepton is obtained from its associated track.
Backgrounds in this measurement are accounted for by subtracting same-sign tag-and-probe pairs from the opposite-sign selection.
 
The templates for the shapes of electron scattering backgrounds are estimated by multiplying the number of events in the single-electron CR by a set of transfer factors.
The electron transfer factors are factorised into two components: pixel tracklet and calorimeter isolation selections, and the final estimate is given by
 
\begin{equation}
f_{e}^{\mathrm{SR}}(\pt; \sigma_{e}, \alpha_{e})= N_{e,\mathrm{signal}}^{\mathrm{CR}} \left( \pt,\eta \right)  \times \mathrm{TF}^{e}_{\textrm{pixel-only}} \left( \pt,\eta \right) \times \mathrm{TF}^{e}_{\textrm{calo-veto}} \left( \pt,\eta \right) \times f(z=z_{\text{ran}}; \sigma_{e}, \alpha_{e})
\label{eqn:eqn_electron_track}
\end{equation}
 
\noindent
where $N_{e,\mathrm{signal}}^{\mathrm{CR}}$ is the number of events in the single-electron CR,
$\mathrm{TF}^{e}_{\textrm{pixel-only}}$ is a transfer factor for an electron failing to satisfy the electron identification and being instead mistakenly categorised as a pixel tracklet,
and $\mathrm{TF}^{e}_{\textrm{calo-veto}}$ is the transfer factor representing the probability for an electron track to be isolated from a calorimeter cluster.
Additionally, the templates are smeared to account for the different momentum resolution of the full-length track and the pixel-tracklets by the $ f(z=z_{\text{ran}}; \sigma_{e}, \alpha_{e})$ term.
The smearing function depends on the electron smearing parameters $\sigma_{e}$ and $\alpha_{e}$ shown in Table~\ref{tab:smearing_parameters},
and is evaluated at a random value ($z_{ran}$) using Equation~\ref{eq:CrystalballLikeFunction}.
 
In the pixel tracklet transfer-factor measurement, the probe lepton is a calorimeter cluster,
and the transfer factor $\mathrm{TF}^{e}_{\textrm{pixel-only}}$ is calculated as the number of probes matched to a pixel tracklet divided by the number of probes matched to an electron.
An object is considered matched if $\Delta R(\text{probe},\text{object})<0.2$.
Both the pixel tracklet and the track matched to the electron must pass the requirements for the pixel tracklet defined in Section~\ref{sec:objects},
without application of the disappearing-track condition for the electron.
The calorimeter isolation transfer factor $\mathrm{TF}^{e}_{\textrm{calo-veto}}$ is calculated using a high-quality inner-detector track as the probe.
The inner-detector track is required to satisfy the same quality criteria as the pixel tracklets, except that instead of meeting the disappearing-track condition, the track is required to have at least eight SCT hits.
This measurement assesses the ratio of the number of inner-detector tracks passing the calorimeter isolation requirement, $E_{\textrm{T}}^{\textrm{topo cluster}} < 5~\GeV$, to the number failing this requirement.
Both transfer factors are parameterised as a function of \pt{} and $\eta$.
Detector simulations are used to correct for the change in the isolation transfer factor when the track selection requires at least eight SCT hits instead of imposing the disappearing-track condition.
These simulation-based corrections can be up to $\mathcal{O}(10^{-4})$ for tracks with \pt{}$>$~30~GeV.
However, since the isolation transfer factor is already very small in this region, these correction factors have a small impact on the resulting template and final results.
 
Since the electron and hadron scattering backgrounds have similar pixel tracklet \pt{} distribution shapes,
the electron transfer factors are used to obtain templates for the hadron scattering backgrounds.
The number of events in a single-inner-detector-track CR is used instead of $N_{e,\mathrm{signal}}^{\mathrm{CR}}$ in Eq.~(\ref{eqn:eqn_electron_track}).
Similarly to the electron $\mathrm{TF}^{e}_{\textrm{calo-veto}}$, detector simulations are used to correct for track-length differences between the two CRs.
Corrections are derived using simulated single-pion events for the hadron scattering backgrounds.
The correction factors can range from approximately 10 to 100, but only have a marginal effect on the pixel tracklet \pt{} template shape.
As the normalization of the hadron scattering backgrounds is extracted from the combined fit described in Section~\ref{sec:background_estimation:fitting},
these correction factors have a small impact on the resulting background prediction and final results.
 
Similarly to the electron measurement, the muon transfer factors are factorised into two components: pixel tracklets and those with no MS track association,
and they are multiplied by the number of events in the single-muon CR to estimate the muon scattering background,
 
\begin{equation}
f_{\mu}^{\mathrm{SR}}(\pt; \sigma_{\mu}, \alpha_{\mu})= N_{\mu,\mathrm{signal}}^{\mathrm{CR}} \left( \pt,\eta,\phi \right)  \times \mathrm{TF}^{\mu}_{\textrm{pixel-only}} \left( \pt,\eta \right) \times \mathrm{TF}^{\mu}_{\textrm{no MS track}} \left( \eta, \phi \right)  \times f(z=z_{\text{ran}}; \sigma_{\mu}, \alpha_{\mu})
\end{equation}
 
where $N_{\mu,\mathrm{signal}}^{\mathrm{CR}}$ is the number of events in the single-muon CR,
$\mathrm{TF}^{\mu}_{\textrm{pixel-only}}$ is the transfer factor representing the probability for a muon to be misidentified as a pixel tracklet,
and $\mathrm{TF}^{\mu}_{\textrm{no MS track}}$ is the transfer factor accounting for the probability that a muon with a good inner-detector track does not have an associated MS track.
Similar to the electron template, the muon template is smeared using the $\sigma_{\mu}$ and $\alpha_{\mu}$ smearing parameters found in Table~\ref{tab:smearing_parameters}.
 
In the muon pixel-tracklet measurement, the probe lepton is a muon track reconstructed only in the muon spectrometer.
The transfer factor $\mathrm{TF}^{\mu}_{\textrm{pixel-only}}$ is calculated in the same way as the electron pixel-tracklet transfer factor, but the denominator is the number of events where the probe is matched to a muon.
The no-MS-track transfer factor $\mathrm{TF}^{\mu}_{\textrm{no MS track}}$ uses a high-quality inner-detector track as the probe.
It represents the probability for an MS track not to be geometrically matched to an inner-detector track.
In order to account for the detector geometry, the transfer factor is measured as a function of $\eta$ and $\phi$.
No significant \pt{} dependence is observed.

\subsection{Combinatorial fake backgrounds\label{sec:background_estimation:fake}}
 
Fake tracklets are reconstructed from a combination of unassociated hits.
These tracklets can have high transverse momentum and can mimic the signal.
Since their $d_0$ is broadly distributed, and the high-$\pt$ chargino tracks have good pointing resolution and originate from the primary interaction region,
the high-$|d_{0}|$ sidebands can be used as a CR that is pure in fake tracklets.
A pure fake-tracklet control sample is obtained by applying the same kinematic selection requirements as in the signal region,
but without the offline $\met$ selection, so as to increase the sample size, and with an inverted requirement of $ |d_0 / \sigma_{d_0}| > 10 $ on the transverse impact parameter.
 
The \pt spectrum of mismeasured pixel tracklets in the fake-tracklet CR is modelled with the following functional form:
 
\begin{equation}
f_c\left(\pt;p_0,p_1\right) = \exp\left(- p_0 \cdot \log(\pt) - p_1 \cdot (\log(\pt))^2\right).
\end{equation}
 
In order to remove any correlations between $p_{0}$ and $p_{1}$ in the fit described in Section~\ref{sec:background_estimation:fitting},
the parameters are rotated by an angle $\theta$ and redefined as $p_{0} = p_{0}{'} \cos\theta - p_{1}{'} \sin\theta$ and $p_{1} = p_{0}{'} \sin\theta + p_{1}{'} \cos\theta$.
The $\theta$ value is chosen such that the two parameters are uncorrelated, and is found to be $\theta=-0.0973$.
The $p_0$ and $p_{1}$ parameters are determined in the combined fit described in Section~\ref{sec:background_estimation:fitting}.
The parameter $p_{1}{'}$ is found to be $p_{1}{'}=0.171 \pm 0.002$ in the signal regions of both channels, and $p_{0}{'} = 1.172 \pm 0.113$ ($0.996 \pm 0.110$) in the electroweak (strong) channel.
The data and fitted fake background are shown in Figure~\ref{figure:bkg:fakes:fitResults} for the electroweak and strong production fake-tracklet CRs and
show good agreement between the fit and the data.
 
\begin{figure}[htbp]
\centering
\subfloat[]{
\includegraphics[width=0.5\textwidth]{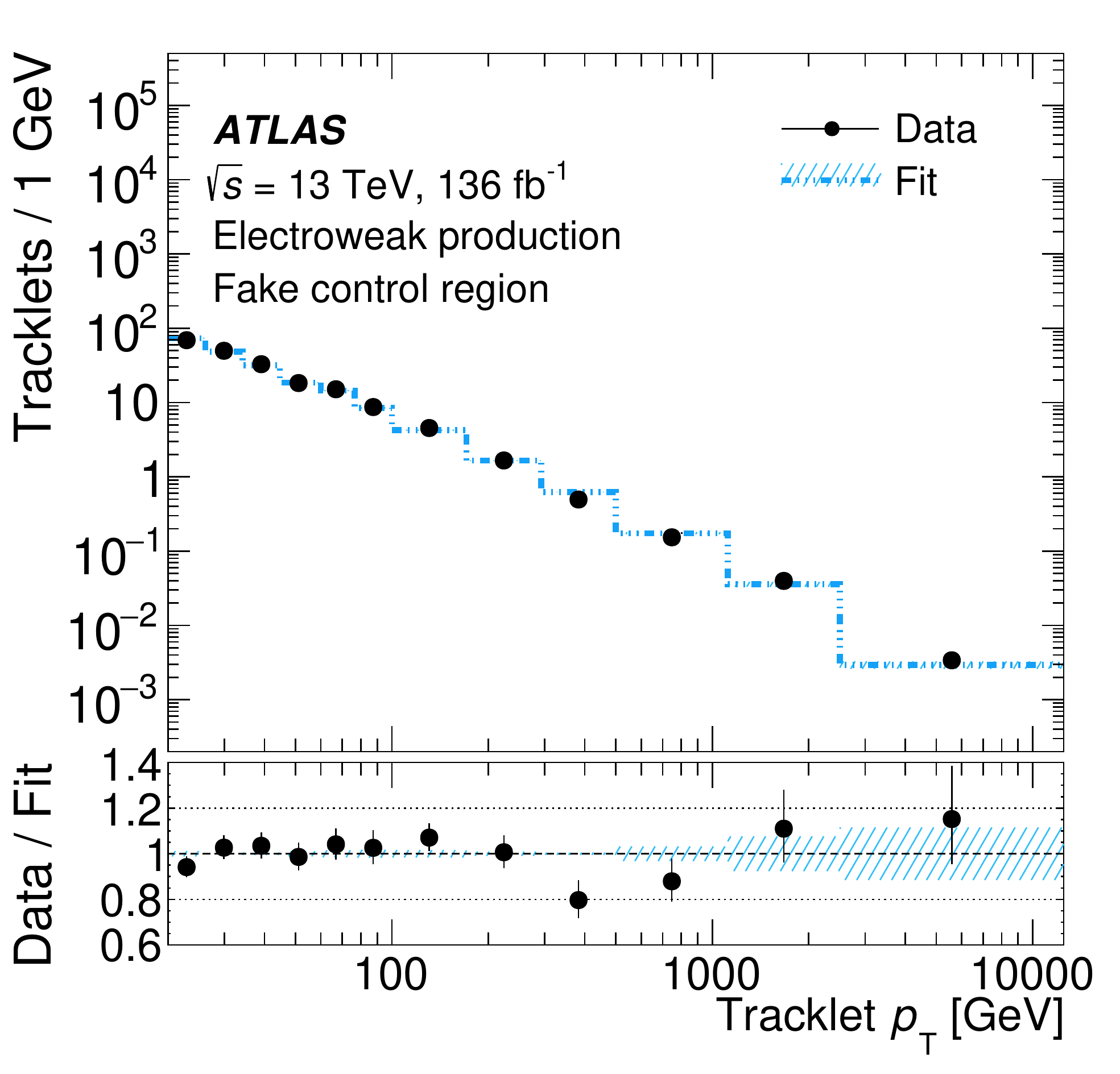}
}
\subfloat[]{
\includegraphics[width=0.5\textwidth]{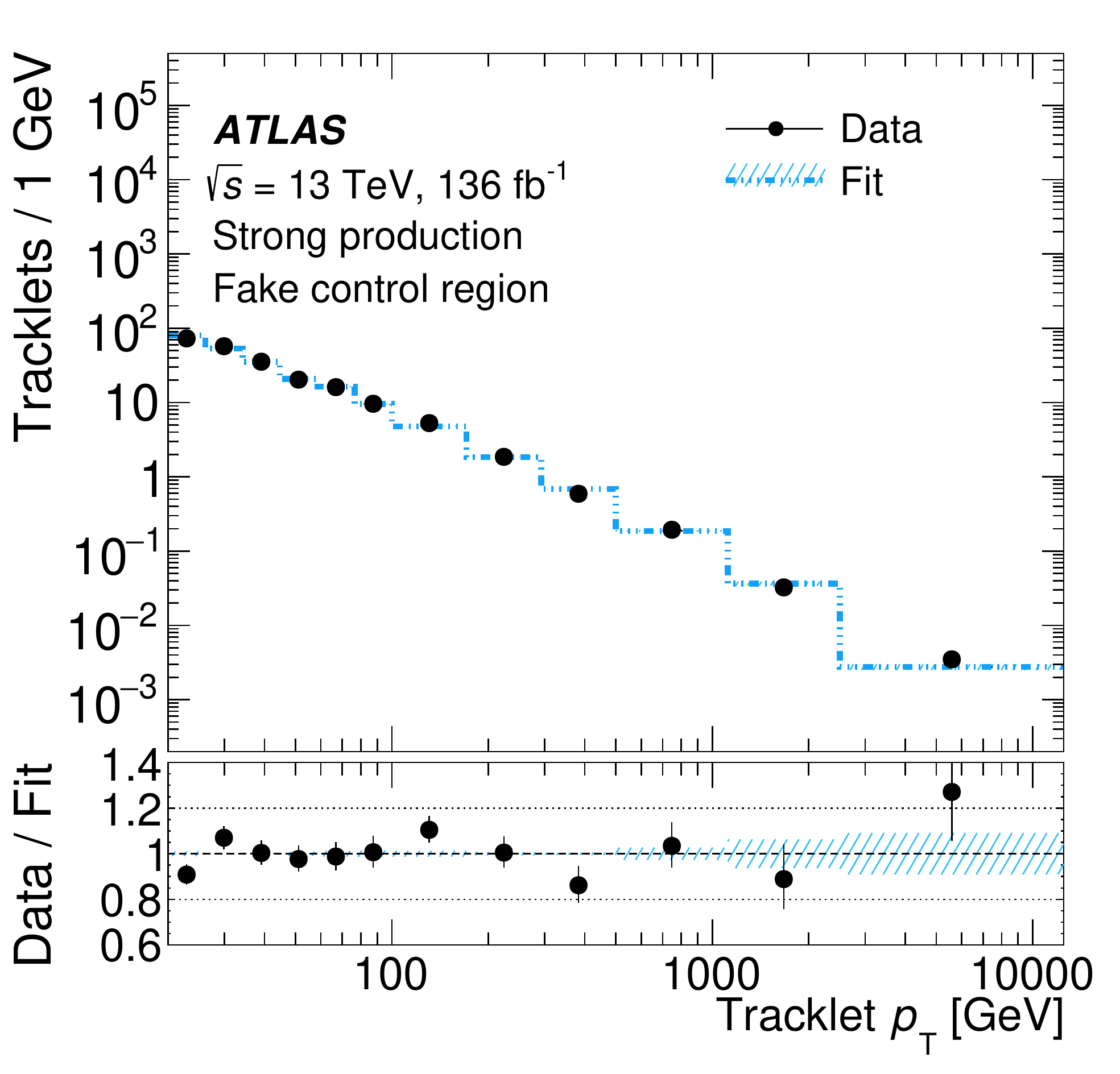}
}
\caption{Fit in the fake-tracklet CR for (a) the electroweak production channel and (b) the strong production channel.
The black points show data.
The blue line and the hatched band show the fit function and its uncertainty.
The bottom insert shows the ratio of the data to the fitted prediction. }
\label{figure:bkg:fakes:fitResults}
\end{figure}

\subsection{Fitting procedure\label{sec:background_estimation:fitting} }
 
The backgrounds are estimated by fitting the model below to the data in the high-\met{} signal region or in each of the middle-\met{} validation regions,
simultaneously with the low-\met{} and fake-tracklet CRs described in Section~\ref{sec:evtSel_CRs}.
The fit is performed using the likelihood function described in the following,
where the likelihood function is described for the high-\met{} signal region setup.
For the validation region fits, the high-\met{} signal region is replaced with the corresponding validation region.
 
The likelihood function $\mathcal{L}$ for the tracklet $\pt$ in a sample of observed events ($N_{\mathrm{obs}}$) is defined as
$\mathcal{L}= {\mathcal{L}_{\mathrm{shape}}} \times {\mathcal{L}_{\mathrm{shape}}^{\mathrm{Fake\,CR}} } \times {\mathcal{L}_{\mathrm{syst}}}$.
The shape terms represent the probability to observe $N_{\mathrm{obs}}$ events in the low-\met{} and high-\met{} region.
The ${\mathcal{L}_{\mathrm{shape}}}$ term is defined as:
 
\begin{equation}
\label{eq:LShape}
\mathcal{L}_{\mathrm{shape}} = \prod_{R}^{\mathrm{low\mathchar`-/high\mathchar`-}\met} \mathcal{L}_{\mathrm{shape}}^{R} ~, \\
\end{equation}
 
\noindent
with
 
\begin{equation}
\mathcal{L}_{\mathrm{shape}}^{R} =
\frac{ \exp\left( -\sum_{i}^{s,e,\mu,h,c} n_{i}^{R} \right) }{{ N_{\mathrm{obs}}^{R} ! }} \times
\prod_{ N_{\mathrm{obs}}^{R} }
\left(
\sum_{i'}^{s,e,\mu,h} \left( n_{i'}^{R} \cdot f_{i'}^{R}(\pt; \sigma_{i'}, \alpha_{i'}) \right)
+ n_{c}^{R} \cdot f_{c}(\pt; p_0, p_1)
\right). \\
\end{equation}
 
The ${\mathcal{L}_{\mathrm{shape}}^{\mathrm{Fake\,CR}} }$ term is defined as:
 
\begin{eqnarray}
\label{eq:LShape_fake}
\mathcal{L}_{\mathrm{shape}}^{\mathrm{Fake\, CR}} &=&
\frac{ \exp{ \left( -n_{c}^{\mathrm{Fake\,CR}} \right) } }{ N_{\mathrm{obs}}^{\mathrm{Fake\,CR} } ! } \times
\prod_{ N_{\mathrm{obs}}^{\mathrm{Fake\,CR}} }
\left(
n_{c}^{\mathrm{Fake\,CR}} \cdot f_c(\pt;p_0,p_1)
\right).
\end{eqnarray}

Here $s, e, \mu, h$ and $c$ are subscripts of the signal process and the electron, muon, hadron and combinatorial fake-tracklet backgrounds respectively;
$N_{\mathrm{obs}}^{R}$ is the number of observed events in each of the fitted regions, $R$ = high-\met{} or low-\met{} region;
$n_{i}^{R}$ is the estimated number of events of process $i$ in region $R$; $\sigma_{i'}$ and $\alpha_{i'}$ are the smearing parameters of process $i'$;
$p_0$ and $p_1$ are the parameters for the combinatorial fake-tracklet background fit described in Section~\ref{sec:background_estimation:fake};
$f_i$ is the charged-particle scattering background-shape template for process $i$ described in Sections~\ref{sec:background_estimation:lepton} and \ref{sec:background_estimation:smearing};
$n_{c}^{R}$ is the estimated number of combinatorial fake-tracklet events in region $R$.
For region R=low-\met{} , the number of combinatorial fake-tracklet events ($n_{c}^{\textrm{low-}\met}$) is a free parameter in the likelihood fit and extracted from the data.
Using this, the number of combinatorial fake-tracklets in the high-\met{} region is calculated from:
 
\begin{equation}
\label{eq:nc}
n_{c}^{\textrm{high-}\met} = r_{\mathrm{CD}} \cdot \exp(r_{\mathrm{ABCD}}) \cdot n_{c}^{\textrm{low-}\met}, \\
\end{equation}
 
\noindent
with $r_{\mathrm{CD}}$ defined as:
 
\begin{equation}
\label{eq:rCD}
r_{\mathrm{CD}}   = \frac{ n_{c}^{\textrm{high-}\met\,\textrm{Fake\,CR }} }{ n_{c}^{\textrm{low-}\met\,\textrm{Fake\,CR}} }, \\
\end{equation}
 
\noindent
and $r_{\mathrm{ABCD}}$ defined as:
 
\begin{equation}
\label{eq:rABCD}
r_{\mathrm{ABCD}} = \ln \frac{ n_{c}^{\textrm{high-}\met} / n_{c}^{{\textrm{high-}\met}\,\textrm{Fake\,CR}} }{ n_{c}^{\textrm{low-}\met} / n_{c}^{{\textrm{low-}\met}\,\textrm{Fake\,CR}} }.
\end{equation}
 
Here $r_{\mathrm{CD}}$ and $r_{\mathrm{ABCD}}$ are terms to constrain the combinatorial fake-tracklet background in the high-\met{} regions relative to that in the low-\met{} regions:
the $r_{\mathrm{CD}}$ term is used for the low-\met{}control and high-\met{} signal region,
while the $r_{\mathrm{ABCD}}$ term is used to constrain the extrapolation from the high $d_{0}$ significance side fake enriched regions into the low-\met{} control and high-\met{} signal regions.
 
The likelihood $\mathcal{L}_{\mathrm{syst}}$ consists of a product of terms related to the systematic uncertainties in each background process and the signal process,
$\mathcal{L}_{\mathrm{syst}} = \mathcal{L}_{\mathrm{syst}}^{s} \times \mathcal{L}_{\mathrm{syst}}^{e} \times \mathcal{L}_{\mathrm{syst}}^{\mu} \times \mathcal{L}_{\mathrm{syst}}^{h} \times \mathcal{L}_{\mathrm{syst}}^{c}$.
Each likelihood component is defined as below:
 
\begin{eqnarray}
\mathcal{L}_{\mathrm{syst}}^{s} &=&
G\left( \sigma_{s}; \overline{\sigma_{s}}, \Delta \sigma_{s} \right) \times
G\left( \alpha_{s}; \overline{\alpha_{s}}, \Delta \alpha_{s} \right) \times
\prod_{R} G\left( n_{s}^{R}; \overline{n_{s}^{R}}, \Delta n_{s}^{R} \right),  \\
\mathcal{L}_{\mathrm{syst}}^{\mu} &=&
G\left( \sigma_{\mu}; \overline{\sigma_{\mu}}, \Delta \sigma_{\mu} \right) \times
G\left( \alpha_{\mu}; \overline{\alpha_{\mu}}, \Delta \alpha_{\mu} \right) \times
\prod_{R} G\left( n_{\mu}^{R}; \overline{n_{\mu}^{R}}, \Delta n_{\mu}^{R} \right),  \\
\mathcal{L}_{\mathrm{syst}}^{e} &=&
G\left( \sigma_{e}; \overline{\sigma_{e}}, \Delta \sigma_{e} \right) \times
G\left( \alpha_{e}; \overline{\alpha_{e}}, \Delta \alpha_{e} \right) \times
\prod_{R} G\left( n_{e}^{R}; \overline{n_{e}^{R}}, \Delta n_{e}^{R} \right),  \\
\mathcal{L}_{\mathrm{syst}}^{h} &=&
G\left( \sigma_{h}; \overline{\sigma_{h}}, \Delta \sigma_{h} \right) \times
G\left( \alpha_{h}; \overline{\alpha_{h}}, \Delta \alpha_{h} \right),  \\
\mathcal{L}_{\mathrm{syst}}^{c} &=&
G\left( r_{\mathrm{ABCD}}; 1, \Delta r_{\mathrm{ABCD}} \right),
\end{eqnarray}
 
where $G(a;b,c)$ represents a unit Gaussian function of $a$ with a mean of $b$ and a standard deviation $c$.
The expected value and the uncertainty of a variable $x$ are represented by $\overline{x}$ and $\Delta x$ respectively.
 
The likelihood is maximised by minimising the negative log-likelihood function with the MINUIT~\cite{James:2004xla} package in the RooFit framework~\cite{RooFit}.
The fit parameters are the normalisations of the hadron and combinatorial fake-tracklet backgrounds ($n_{h}^{R}$ and $n_{c}^{R}$),
the $p_0$ and $p_1$ parameters of the function modelling the fake-tracklet transverse momentum distribution described in Section~\ref{sec:background_estimation:fake}, and nuisance parameters.
Each nuisance parameter represents a source of systematic uncertainty and is allowed to float in the fit with a Gaussian constraint.
The statistical uncertainty of the transfer factors for electrons and muons is propagated into the final template.

% End of text imported from the .//sections/background_estimation.tex input file
% The next lines are included from the .//sections/systematics.tex input file
 
\section{Systematic uncertainties\label{sec:syst}}
 
Several sources of uncertainty are considered for this search,
and these can be divided into uncertainties affecting the parameters in the background and signal fit model,
and those affecting the expected signal yields.
 
\subsection{Fit model uncertainties}
 
Uncertainties in the normalisation of the electron and muon backgrounds are dominated by the statistical uncertainties in the transfer factors.
The shape uncertainty in the hadron and charged-lepton backgrounds is dominated by uncertainties in the smearing functions.
The pile-up condition is the largest source of uncertainty for the smearing function, especially for lower-\pT{} tracklets.
The pile-up uncertainty is evaluated by taking the difference between the nominal values of the smearing parameters and the values obtained from events with different pile-up conditions.
This is done by splitting the dataset into data taken in low ($\mu<40$) and high ($\mu>40$) pile-up conditions, and the full parameter-value differences between the datasets with low and high pile-up are assigned as the systematic uncertainties of the \pt{}-dependent smearing parameters.

The uncertainties in $r_{\textrm{CD}}$ and $r_{\textrm{ABCD}}$ are obtained from statistical uncertainties in the control regions used to calculate these parameters.
An additional uncertainty in the extrapolation of $r_{\textrm{ABCD}}$ over $d_0/\sigma(d_0)$ is obtained by
evaluating $r_{\textrm{ABCD}}$ using pixel tracklets with $3<|d_0/\sigma(d_0)|<10$ for $n_{c}^{\textrm{high-}\met}$ and  $n_{c}^{\textrm{low-}\met}$ in Eq.~(\ref{eq:rABCD}).
Since fake tracklets are the dominant backgrounds in the signal regions, variations of these parameters are the leading source of uncertainty in both the electroweak and strong production signal regions.
Similarly, the $p_{0}$ and $p_{1}$ parameters are varied up and down by the statistical uncertainties obtained from the fit.
 
Table~\ref{table:background_uncertainty1} summarises the effect of various sources of systematic uncertainty on the signal exclusion significance.
 
\begin{table}[thb]
\centering
\begin{footnotesize}
\caption{Effects of systematic uncertainties on the signal exclusion significance for a representative signal point with $m_{\chinoonepm} = 600~\GeV{}$
for the electroweak channel and $m_{\tilde{g}} = 1400~\GeV{}$ and $m_{\chinoonepm} = 1100~\GeV{}$ for the strong channel.
}
\label{table:background_uncertainty1}
\begin{tabular}{lcc} 
\toprule
& {Electroweak channel~[\%]} & {Strong channel~[\%]} \\
\midrule
$r_{\textrm{ABCD}}$                                     &    5.2 &    0.9 \\
$r_{\textrm{CD}}$                                       &    3.2 &    0.6 \\
$\sigma$ parameter in signal \pT{} smearing function              &    2.9 &    0.1 \\
$\alpha$ parameter in signal \pT{} smearing function              &    1.7 &    0.2 \\
$p_{0}$ parameter in the fake background \pT{} function &    0.3 & $<$0.1 \\
$p_{1}$ parameter in the fake background \pT{} function &    0.3 &    0.2 \\
Normalisation of muon background                        &    0.6 & $<$0.1 \\
Normalisation of electron background                    & $<$0.1 & $<$0.1 \\
$\alpha$ parameter in muon \pT{} smearing function                & $<$0.1 & $<$0.1 \\
$\sigma$ parameter in muon \pT{} smearing function                & $<$0.1 & $<$0.1 \\
$\alpha$ parameter in electron \pT{} smearing function            & $<$0.1 & $<$0.1 \\
$\sigma$ parameter in electron \pT{} smearing function            & $<$0.1 & $<$0.1 \\
$\alpha$ parameter in hadron \pT{} smearing function              &    0.5 &    0.2 \\
$\sigma$ parameter in hadron \pT{} smearing function              &    0.6 &    0.2 \\
\bottomrule
\end{tabular}
\end{footnotesize}
\end{table}

\subsection{Signal uncertainties}
 
A breakdown of the systematic uncertainties for the expected number of signal events in the signal regions is shown in Table~\ref{table:signal_uncertainty1}.
 
Theoretical uncertainties in the signal cross-section are estimated in a way similar to that for the previous result\,\cite{SUSY-2016-06}.
This is done by computing the changes in the cross-section when the renormalisation and factorisation scales, the choice of PDFs and the strong coupling constant, $\alphas$, are varied independently.
Renormalisation and factorisation scales are varied by factors of 0.5 and 2 from their nominal value.
The PDF uncertainty is estimated as the maximum of the uncertainty from the CTEQ6.6\,\cite{Pumplin:2002vw} uncertainty band at
68\% confidence level and the difference between the results for the CTEQ6.6 and MSTW2008 NLO PDF~\cite{Martin:2009iq} sets.
Each uncertainty is varied independently and their effects are added in quadrature.
Uncertainties in the modelling of ISR and final-state radiation (FSR) are estimated by varying the renormalisation, factorisation and merging scales from 0.5 to 2 times their nominal values.
Jet modelling uncertainties related to the energy scale and resolution, and jet vertex tagging uncertainty, are estimated by comparing simulated events and data as described in Ref.~\cite{JETM-2018-05}.
The pile-up modelling uncertainty is estimated by varying the number of collisions per bunch crossing in simulation by its uncertainty of 10\% of the nominal value.
The uncertainty in the \met{} soft-term modelling is considered by comparing data and simulated samples using $Z\rightarrow \mu\mu$ + jets events as described in Ref.~\cite{JETM-2018-05}.
The uncertainty in the trigger efficiency modelling is small because it is measured from data as described in Section~\ref{sec:event_selection}.
Only the statistical uncertainty in the efficiency measurement is taken into account for the signal trigger efficiency uncertainty.
 
The uncertainty in the tracklet reconstruction efficiency is evaluated by comparing efficiencies for muons in the collision data and simulation. Events containing muons from $Z$ boson decays are selected and muons are re-tracked using only pixel hits as described in Section~\ref{sec:background_estimation:smearing}. The efficiency is measured as the ratio of re-tracked pixel tracklets satisfying all of the tracklet quality and selection criteria to the total number of muon tracks. The difference between data and simulation is assigned as a systematic uncertainty.
 
\begin{table}[thb]
\centering
\begin{footnotesize}
\caption{
Effects of systematic uncertainties on the signal yields for representative signal points with $\tau_{\chinoonepm}$~=~0.2~ns.
Each value is obtained by symmetrising the larger of the upward and downward deviations from the nominal signal yield.
The effects of pixel tracklet efficiency and luminosity uncertainties are common to the two signal regions.
}
\label{table:signal_uncertainty1}
\begin{tabular}{lcc}
\toprule
& {Electroweak channel~[\%]} & {Strong channel~[\%]} \\
& {$m_{\chinoonepm} = 600~\GeV$} & {$m_{\gluino}=1400~\GeV$} \\
&                                & {$m_{\chinoonepm}= 1100~\GeV$} \\
\midrule
Cross-section   &   7.6 &  14\\
\midrule
Initial/final-state radiation      &    8.4 &   5.1 \\
Jet energy scale                   &    2.3 &   1.5 \\
Jet energy resolution              &    0.6 &   0.3 \\
Jet vertex tagging efficiency      &   $<$0.1 &  $<$0.1 \\
Pile-up modelling                  &    0.7 &  $<$0.1 \\
\met{} soft term                   &    0.4 &  $<$0.1 \\
Trigger efficiency                 &    0.3 &   0.4 \\
Tracklet reconstruction efficiency                &  \multicolumn{2}{c}{5.9} \\
Luminosity                         &  \multicolumn{2}{c}{1.7} \\
\midrule
Total                   &  11 &  8.1 \\
\bottomrule
\end{tabular}
\end{footnotesize}
\end{table}
 
% End of text imported from the .//sections/systematics.tex input file
% The next lines are included from the .//sections/results.tex input file
 
\section{Results and interpretation\label{sec:results}}
 
The transverse momentum spectra of the fitted pixel tracklets in the low-\met{} CR, the calorimeter sideband and middle-\met{} VRs, and the high-\met{} SR for the electroweak and strong production channels
are shown in Figures~\ref{figure:results:fit_SRs_EWK} and~\ref{figure:results:fit_SRs_Strong}, respectively.
The measured spectra show agreement with the sum of all expected background.
Examples of the expected-signal prediction with $m_{\chinoonepm}=600$~\GeV{} and $\chinoonelife~=~0.2$~ns
for the electroweak production channel, and with $m_{\gluino}=1400$~\GeV, $m_{\chinoonepm}=1100$~\GeV{} and $\chinoonelife~=~0.2$~ns for the strong production channel are overlaid.
 
The numbers of events in the two middle-$\met$ validation regions after the background fits described in Section~\ref{sec:background_estimation:fitting} are shown in Table~\ref{table:summary_vr}.
Good agreement between data and background predictions is observed in all validation regions.
No significant excess above the background predictions is observed in the high-\met{} signal regions, as shown in Table~\ref{table:summary_bg}.
The probability of a background-only experiment being more signal-like than observed ($p_0$), its equivalent formulation in terms of the number of standard deviations ($Z$),
and the upper limit on the model-independent visible cross-section at 95\% confidence level (CL) using the CL$_\textrm{s}$ technique~\cite{Read:2002hq} are also shown in Table~\ref{table:summary_bg}.
For the evaluation of these model-independent quantities, the tracklet is required to have $\pt{}>60$~GeV.
 
\begin{table}[htbp]
\centering
\caption{Expected and observed numbers of events and corresponding background predictions in the middle-$\met$ validation regions.
The uncertainty in the total background prediction differs from the sum in quadrature of the individual components due to anti-correlation of fit parameters between the backgrounds.
}
\label{table:summary_vr}
\begin{tabular}{ccccc}
\toprule
& \multicolumn{2}{c}{Electroweak channel} & \multicolumn{2}{c}{Strong channel} \\
\midrule
& {Calo.\ sideband}  & {Low $\pt$}        & {Calo.\ sideband}   & {Low $\pt$} \\
& {$\pt > 60 ~\GeV$}& {$\pt < 60 ~\GeV$} & {$\pt > 60 ~\GeV$} & {$\pt < 60 ~\GeV$} \\
\midrule

Fake              & 4.3    $\pm$ 2.2     &  5.5   $\pm$ 1.5   & 3.2    $\pm$ 1.5    &  3.5    $\pm$ 1.0   \\
Hadron            & 1.0    $\pm$ 0.8     & 23     $\pm$ 6~~     & 0.36   $\pm$ 0.23   & 13      $\pm$ 4~~     \\
Electron          & 0.8    $\pm$ 0.5     &  1.2   $\pm$ 1.3   & 0.29   $\pm$ 0.20   &  0.5    $\pm$ 0.5   \\
Muon              & 0.023  $\pm$ 0.007   &  0.25 $\pm$ 0.06   & 0.012 $\pm$ 0.004   &  0.129  $\pm$ 0.032 \\
 
\midrule
Total Expected    & 6.1   $\pm$ 1.9      & 29  $\pm$ 5~~     & 3.8   $\pm$ 1.5      & 17   $\pm$ 4~~ \\
\midrule
Observed     & \multicolumn{1}{c}{5} & \multicolumn{1}{c}{30} & \multicolumn{1}{c}{3} & \multicolumn{1}{c}{18} \\
\bottomrule
\end{tabular}
\end{table}
 
\begin{table}[htbp]
\centering
\caption{Expected and observed numbers of events and corresponding background predictions in the high-$\met$ signal regions with transverse momenta above 60~\GeV.
The uncertainty in the total background prediction differs from the sum in quadrature of the individual components due to anti-correlation of fit parameters between the backgrounds.
}
\begin{tabular}{ccc}
\toprule
& {Electroweak channel} & {Strong channel} \\
\midrule
Fake      & 2.6  $\pm$ 0.8      & 0.77   $\pm$ 0.33 \\
Hadron    & 0.26  $\pm$ 0.13    & 0.024  $\pm$ 0.031 \\
Electron  & 0.021 $\pm$ 0.023   & 0.004  $\pm$ 0.004 \\
Muon      & 0.17 $\pm$ 0.06     & 0.049  $\pm$ 0.018 \\
\midrule
Total Expected    & 3.0 $\pm$ 0.7   & 0.84 $\pm$ 0.33 \\
\midrule
Observed     & \multicolumn{1}{c}{3} & \multicolumn{1}{c}{1} \\
\midrule
$p_0$ ($Z$)  & {0.5 ~(0)} & {0.38 ~(0.30)} \\
Observed $\sigma_{\mathrm{vis}^{95\%}}$ [fb] &  {0.037} &  {0.028} \\
Expected $\sigma_{\mathrm{vis}^{95\%}}$ [fb] &  {0.038 $_{-0.009}^{+0.014}$} & {0.024 $_{-0.003}^{+0.009}$} \\
\bottomrule
\end{tabular}
\label{table:summary_bg}
\end{table}

\begin{figure}[htbp]
\centering
\subfloat[]{ \includegraphics[width=0.48\textwidth]{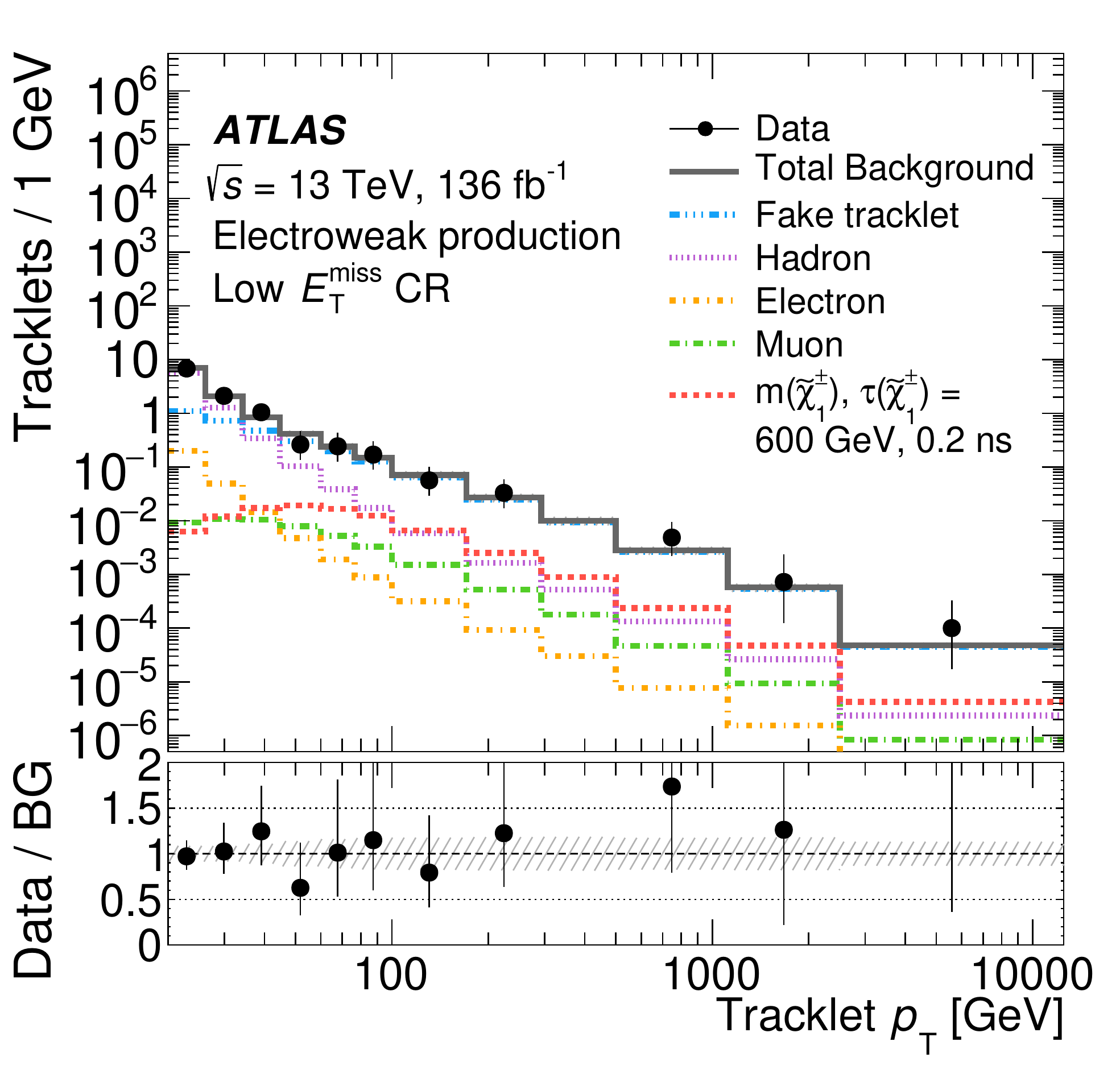}}
\subfloat[]{ \includegraphics[width=0.48\textwidth]{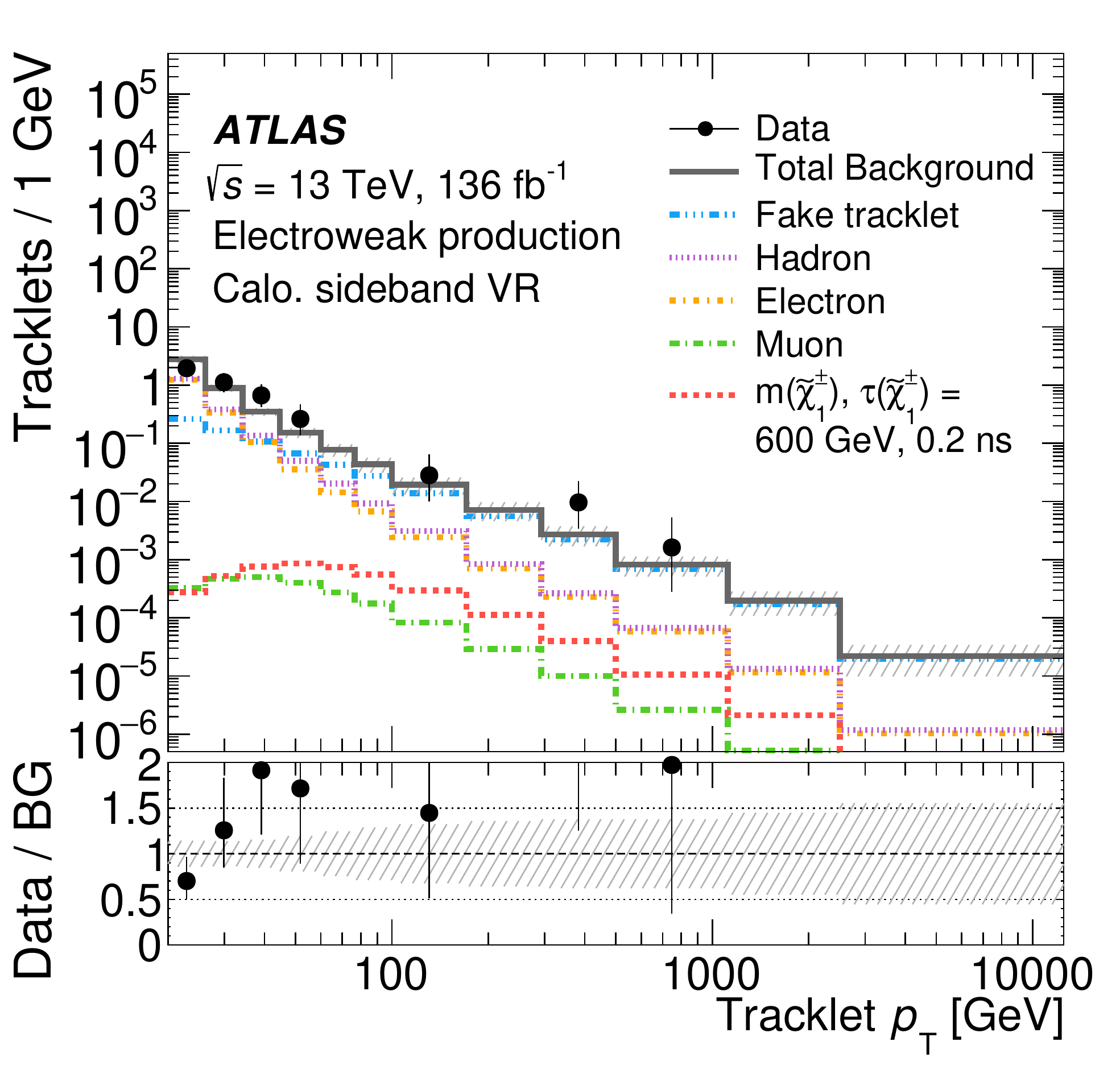}} \\
\subfloat[]{ \includegraphics[width=0.48\textwidth]{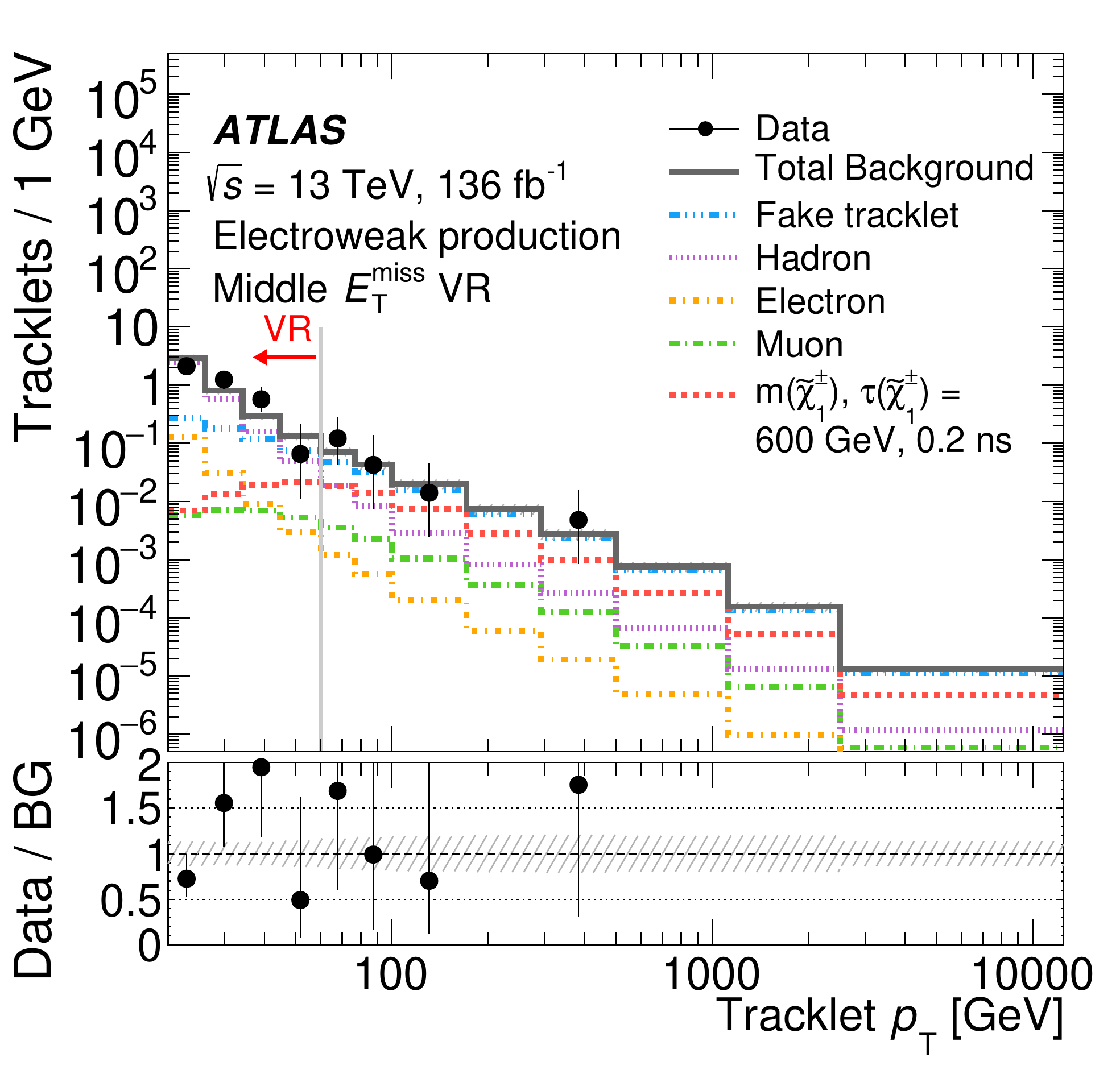}}
\subfloat[]{ \includegraphics[width=0.48\textwidth]{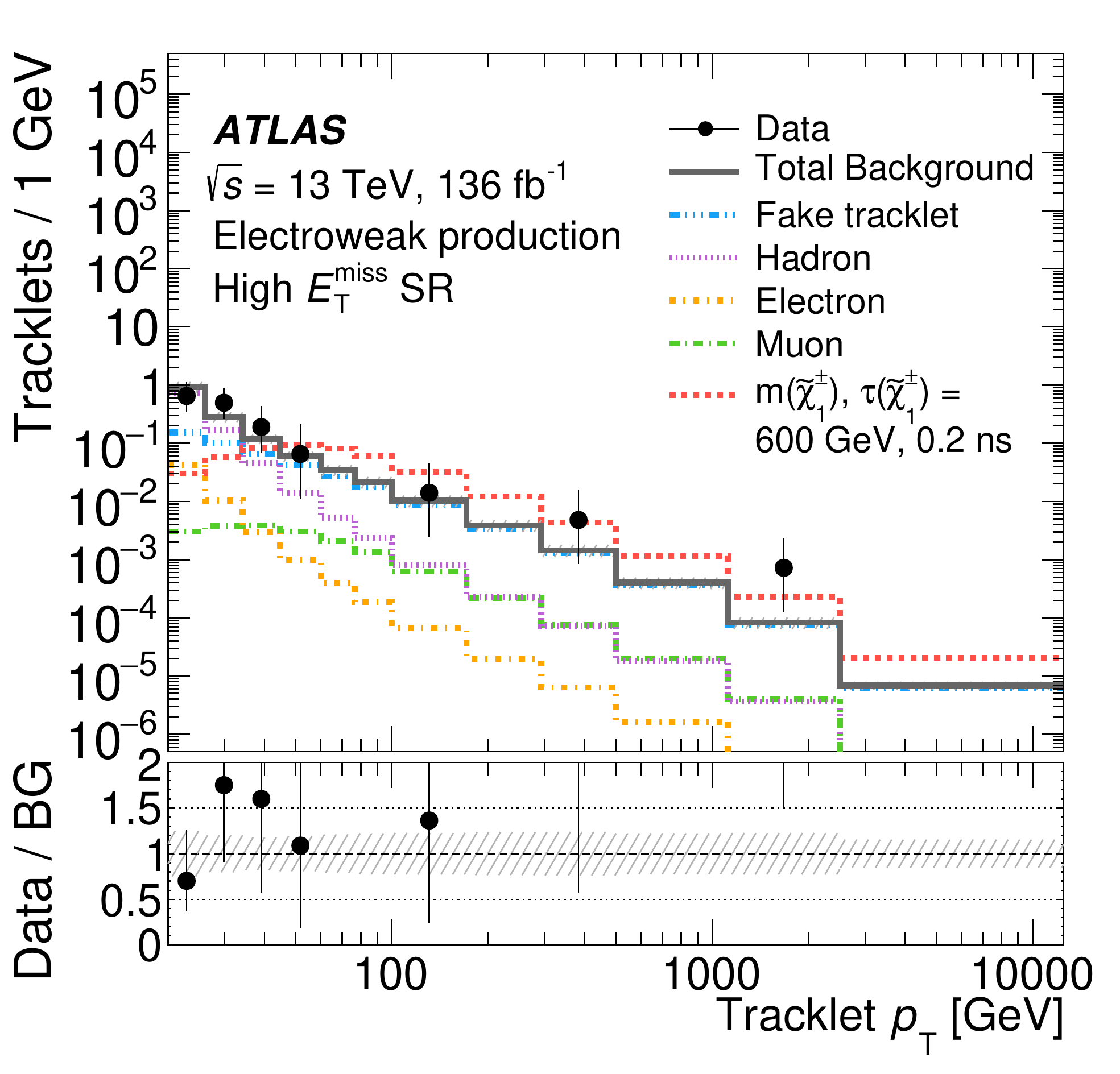}}
\caption{Pixel tracklet \pt{} spectrum fit results in the low-, middle- and high-\met{} regions for the electroweak production channel.
The fake-tracklet, hadron, electron and muon templates are fitted to observed data events in a background-only fit.
An example of the expected-signal prediction with $m_{\chinoonepm}=600$~\GeV{} and $\chinoonelife~=~0.2$~ns is overlaid in red.
The last bin includes overflow entries.
The bottom panel shows the ratio of data to the background-only prediction.
}
\label{figure:results:fit_SRs_EWK}
\end{figure}
 
\begin{figure}[htbp]
\centering
\subfloat[]{ \includegraphics[width=0.48\textwidth]{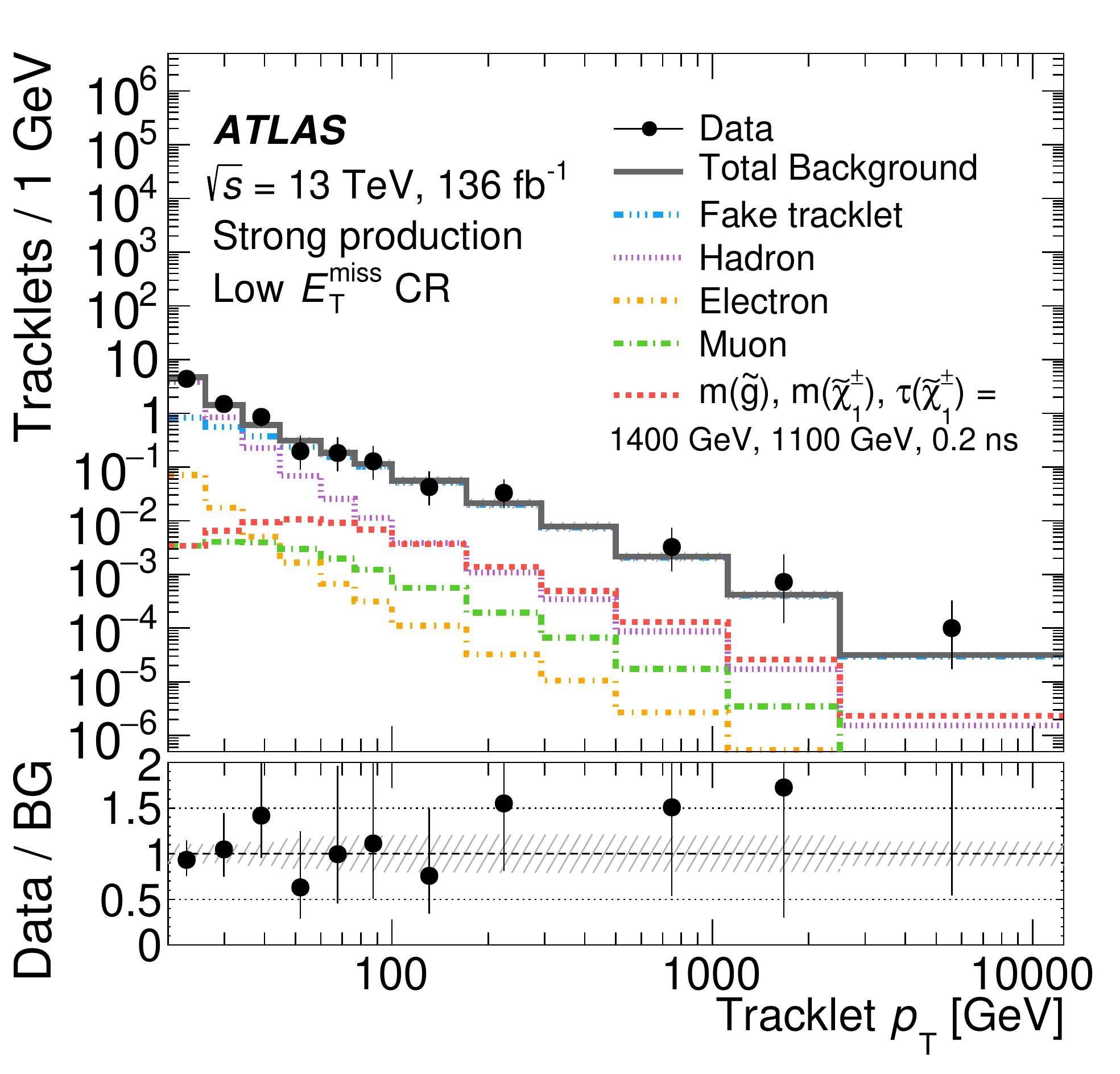}}
\subfloat[]{ \includegraphics[width=0.48\textwidth]{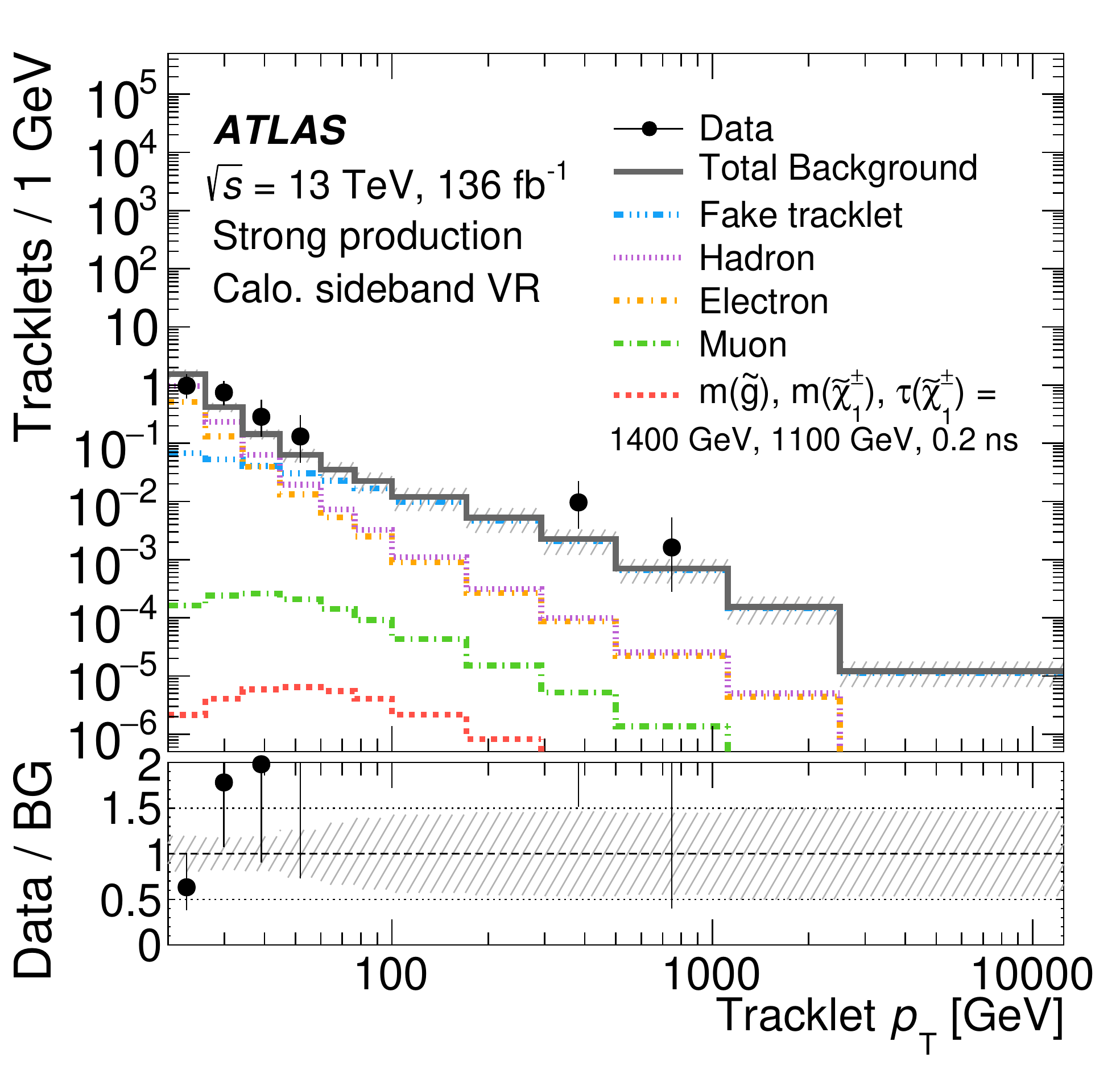}} \\
\subfloat[]{ \includegraphics[width=0.48\textwidth]{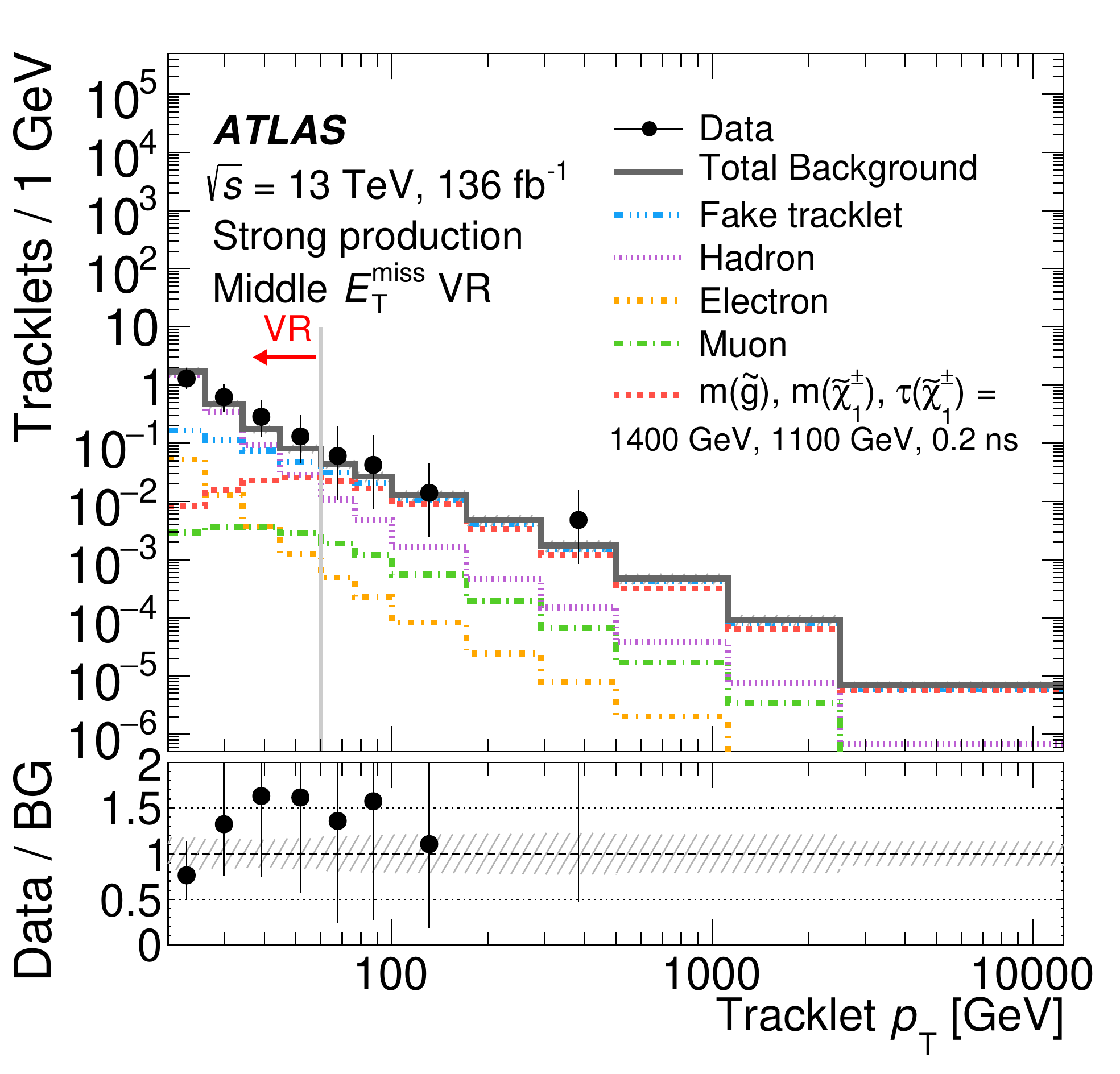}}
\subfloat[]{ \includegraphics[width=0.48\textwidth]{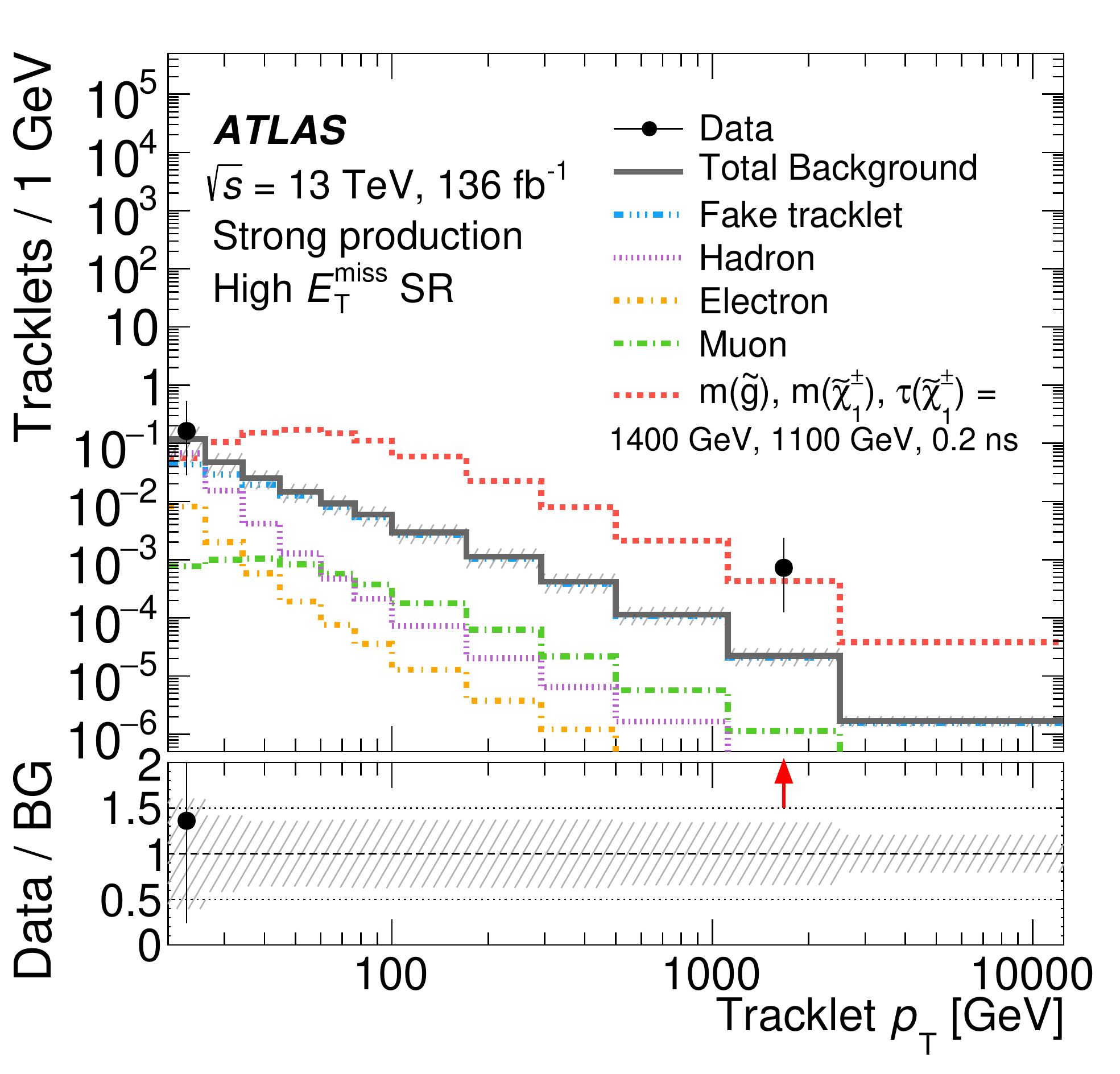}}
\caption{Pixel tracklet \pt{} spectrum fit results in the low-, middle- and high-\met{} regions for the strong production channel.
The fake-tracklet, hadron, electron and muon templates are fitted to observed data events in a background-only fit.
An example of the expected-signal prediction with $m_{\gluino}=1400$~\GeV, $m_{\chinoonepm}=1100$~\GeV{} and $\chinoonelife~=~0.2$~ns
is overlaid in red.
The last bin includes overflow entries.
The bottom panel shows the ratio of data to the background-only prediction.
The arrow indicates a value outside of the range of the bottom inset.
}
\label{figure:results:fit_SRs_Strong}
\end{figure}
 
Model-dependent exclusion limits at 95\% CL are placed on the various signal models.
The likelihood function is extended to contain both the signal and background components.
The signal normalisation is the parameter of interest and is unconstrained in the fit.
 
For the electroweak production of pure winos or pure higgsinos, the exclusion limits are shown as a function of the chargino lifetime and mass.
 
In the case of pure winos, the expected lower limit on the mass of a chargino with a lifetime of 0.2~ns is improved from 450~$\gev$\ in the previous search~\cite{SUSY-2016-06} to 680~$\gev$\ in this search, while a simple scaling from the integrated luminosity used in the previous search gives an expected mass limit of 530~$\gev$\ with the present integrated luminosity of 136~\ifb. The observed limit excludes chargino masses up to 660~\GeV\ as shown in Figure~\ref{figure:ExpectedLimit_ewk}, in which the pure-wino lifetime is shown by the grey dashed line.
 
For pure higgsinos, chargino masses are excluded up to 210~\GeV\ as shown in Figure~\ref{figure:ExpectedLimit_higgsino}.
The lower sensitivity to higgsinos in comparison with winos is explained by two factors:
the smaller production cross-section of higgsinos relative to the wino case,
and the extremely short lifetime of the higgsinos.
 
For the strong production channels, the exclusion limits are set as a function of the lightest chargino's mass and the gluino's mass.
The exclusion limits for charginos with lifetimes fixed to 0.2~ns and 1.0~ns are shown in Figure~\ref{figure:ExpectedLimit_strong}.
The observed limits from the electroweak wino model in Figure~\ref{figure:ExpectedLimit_ewk} are shown by a green hatched line.
Gluino masses below 2.1~\TeV{} are excluded for a chargino mass of 300~\gev{} and a chargino lifetime of 0.2~ns.

\begin{figure}[htbp]
\centering
\includegraphics[width=0.8\textwidth]{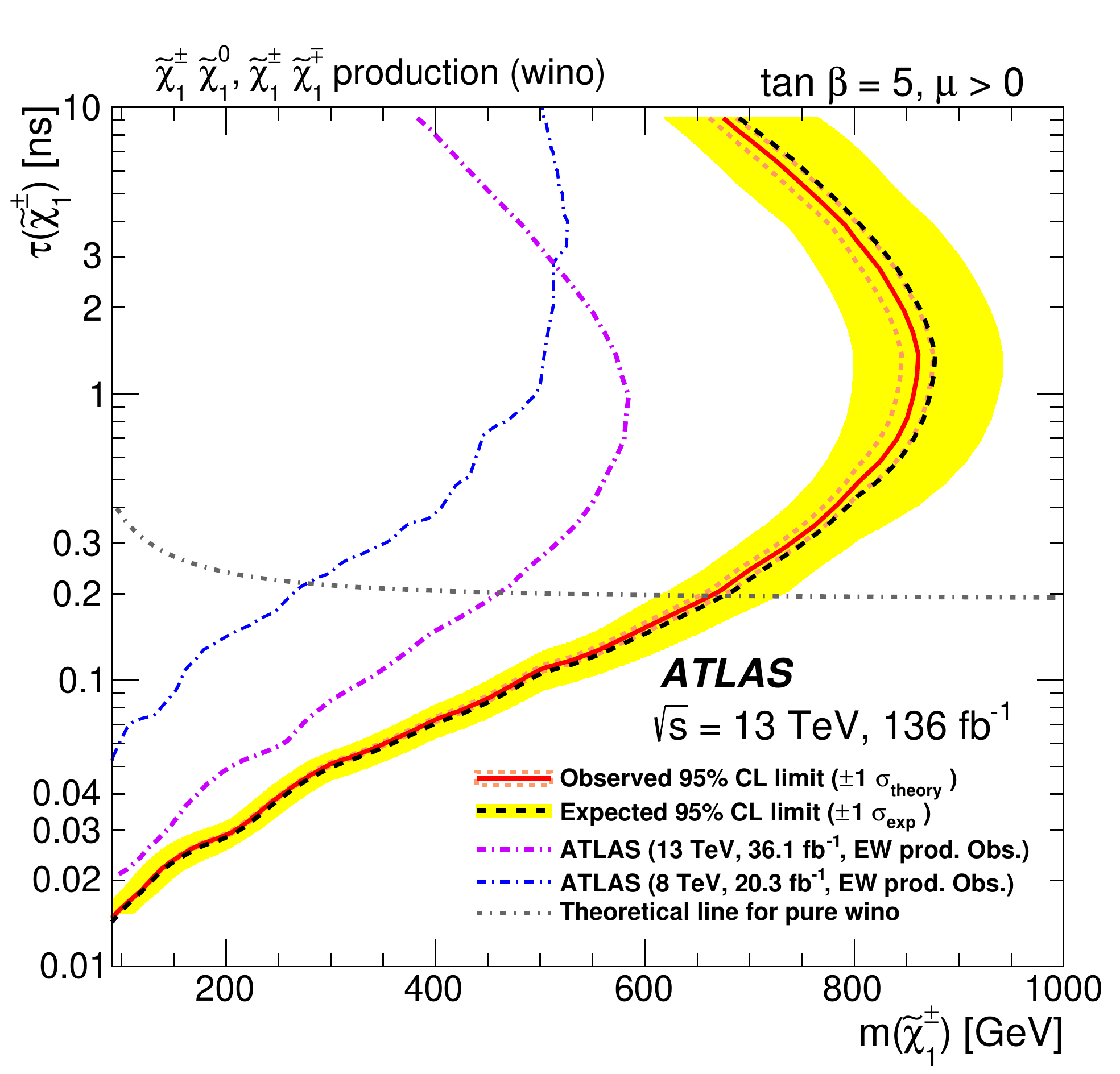}
\caption{Exclusion limits at 95\% CL obtained in the electroweak production channel with the pure-wino scenario.
The limits are shown as a function of the chargino lifetime and mass.
The black dashed line shows the median expected value, and the yellow band shows the 1$\sigma$ uncertainty band around the expected limits.
The red line shows the observed limits and the red dotted lines show the 1$\sigma$ uncertainty from the signal cross-section.
The blue and violet broken lines show the observed limits from the ATLAS results in Refs.~\cite{SUSY-2016-06} and \cite{SUSY-2013-01} respectively.
The dashed gray line shows the predicted chargino lifetime in the almost pure wino-LSP scenario at the two-loop level~\cite{Ibe:2012sx}.
}
\label{figure:ExpectedLimit_ewk}
\end{figure}
 
\begin{figure}[htbp]
\centering
\subfloat[]{
\includegraphics[width=0.5\textwidth]{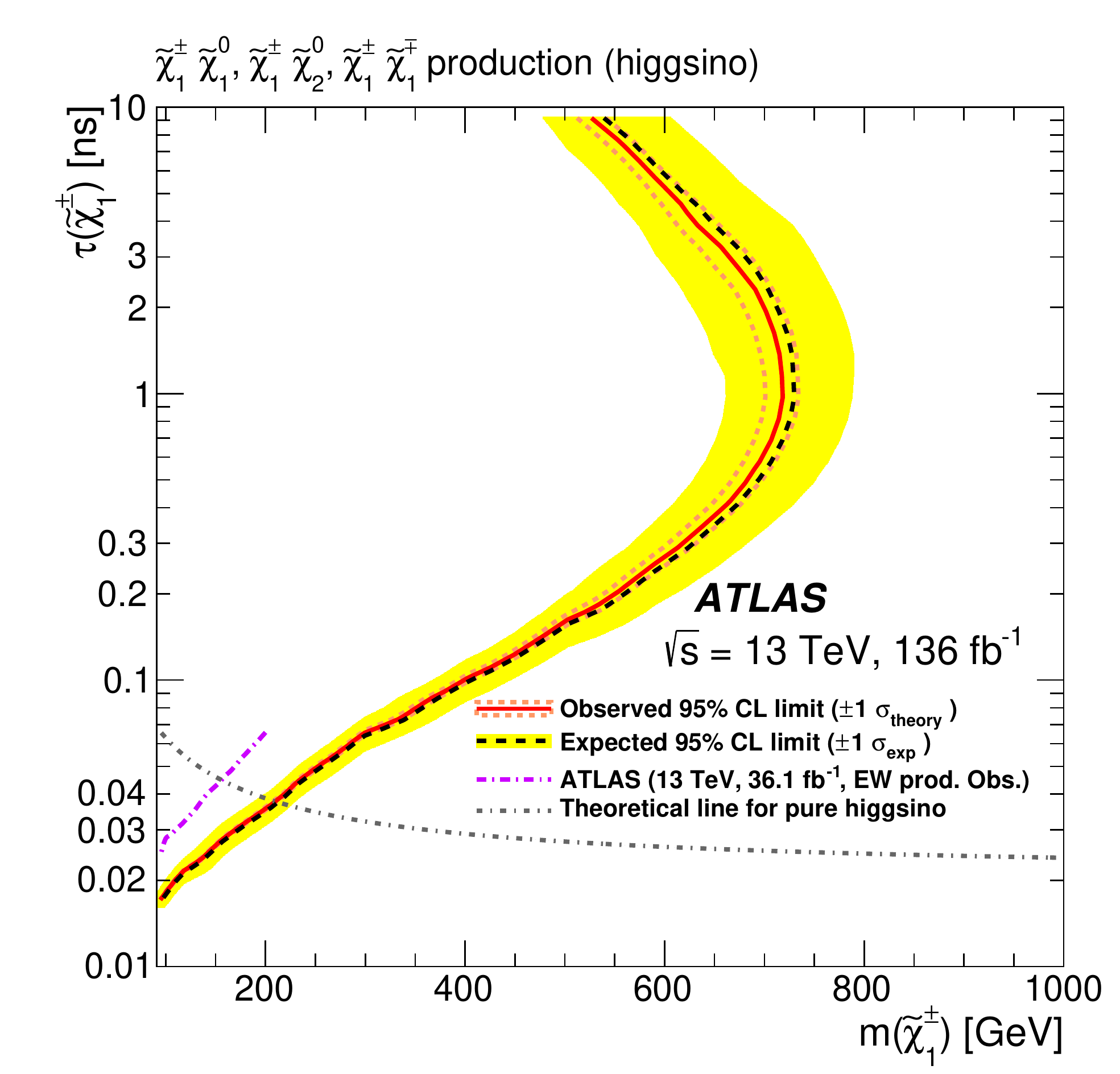}
}
\subfloat[]{
\includegraphics[width=0.5\textwidth]{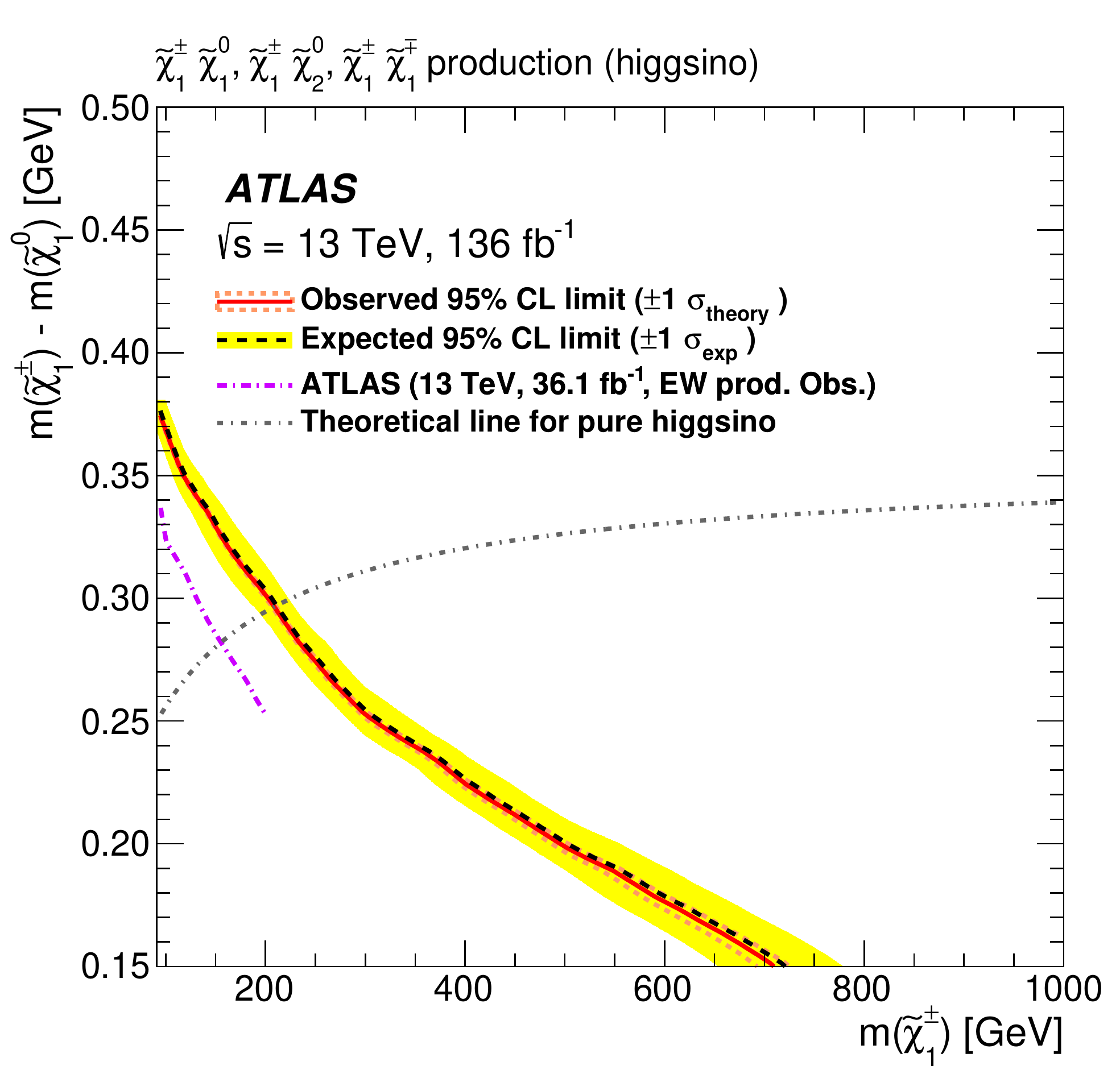}
}
\caption{Exclusion limits at 95\% CL obtained in the electroweak production channel with the pure-higgsino scenario.
The limits are shown separately for the higgsino lifetime or mass splitting as a function of the chargino mass.
The black dashed line shows the median expected value, and the yellow band shows the 1$\sigma$ uncertainty band around the expected limits.
The red line shows the observed limits and the red dotted lines show the 1$\sigma$ uncertainty from the signal cross-section.
The violet broken line shows the observed limits from the previous ATLAS result~\cite{ATL-PHYS-PUB-2017-019}.
The dashed gray line shows the predicted chargino lifetime in the pure higgsino-LSP scenario~\cite{PhysRevLett.81.34}.}
\label{figure:ExpectedLimit_higgsino}
\end{figure}

\begin{figure}[htbp]
\centering
\subfloat[]{
\includegraphics[width=0.5\textwidth]{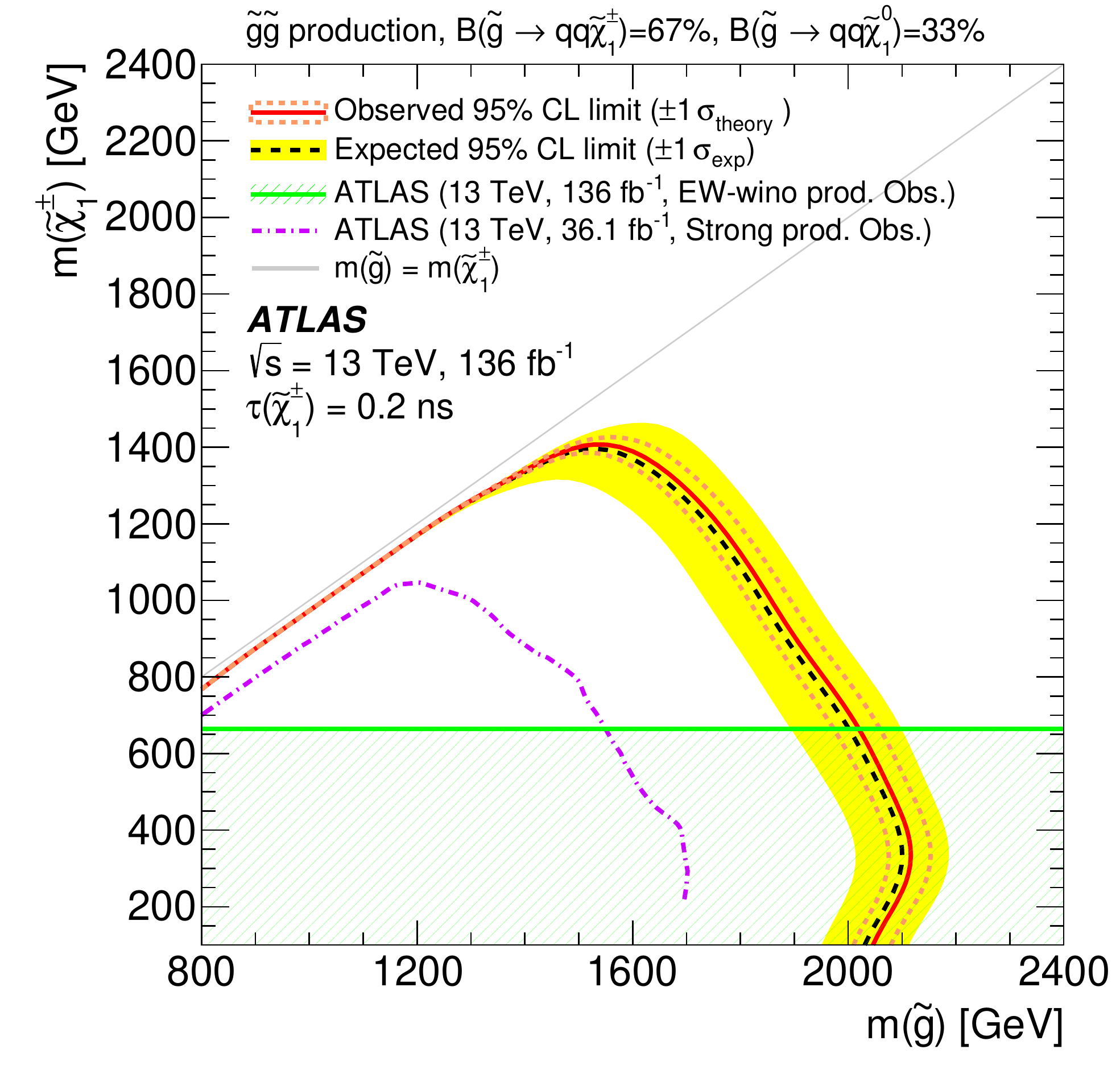}
}
\subfloat[]{
\includegraphics[width=0.5\textwidth]{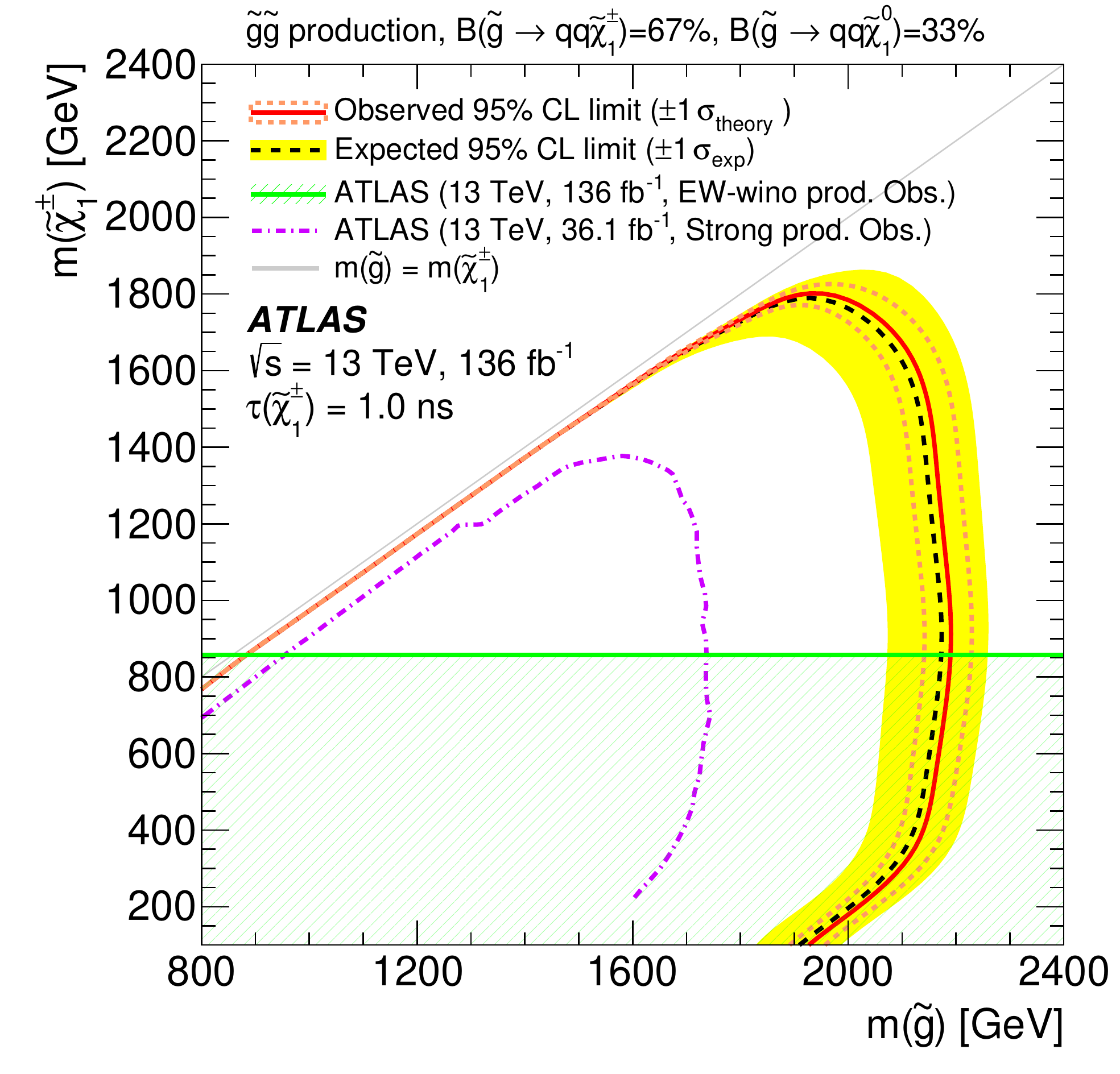}
}
\caption{Exclusion limits at 95\% CL obtained in the strong production channel.
The limits are shown as a function of the lightest chargino's mass versus the gluino's mass, for chargino lifetimes of (a) 0.2~ns and (b) 1.0~ns.
The black dashed line shows the median expected value, and the yellow band shows the 1$\sigma$ uncertainty around the expected limits.
The red line shows the observed limits and the red dotted lines show the 1$\sigma$ uncertainty from the signal cross-section.
The observed exclusion limits from this search in the direct electroweak wino production channel in the minimal AMSB model with
(a) $\tau_{\chinoonepm}$~=~0.2~ns and (b) $\tau_{\chinoonepm}$~=1.0~ns are overlaid in green.
The violet broken line shows the observed limits from the previous ATLAS result~\cite{SUSY-2016-06}.}
\label{figure:ExpectedLimit_strong}
\end{figure}
% End of text imported from the .//sections/results.tex input file
\FloatBarrier
% The next lines are included from the .//sections/conclusion.tex input file
 
\section{Conclusion\label{sec:conclusion}}
 
A search for long-lived charginos with a disappearing-track signature was performed using
$\sqrt{s} = 13~\TeV{}$ $pp$ collision data collected by the ATLAS experiment at the LHC, corresponding to an integrated luminosity of 136~\ifb{}.
Pixel tracklets with at least four hits in the pixel detector are used to improve the sensitivity for short chargino lifetimes.
A strong disappearing-track condition, enforced by vetoes on SCT hits and significant calorimeter energy along the tracklet's trajectory,
was developed specifically for this search and is used to reduce the backgrounds significantly.
The improved background rejection that this disappearing-track condition offers, together with over a factor of three increase in integrated luminosity,
has allowed the ATLAS Collaboration to significantly improve sensitivity to long-lived charginos.
A lower limit on the mass of long-lived charginos from electroweak production in pure-wino (pure-higgsino) models is set at 660~(210)~\GeV{} at 95\% CL.
If charginos with a proper lifetime of 0.2 ns are produced in the decay cascade of pair-produced gluinos, gluino masses below 2.1~\TeV{} are excluded for a chargino mass of 300~\GeV{}.
% End of text imported from the .//sections/conclusion.tex input file

\section*{Acknowledgements}
 
% The next lines are included from the .//acknowledgements/Acknowledgements.tex input file

We thank CERN for the very successful operation of the LHC, as well as the
support staff from our institutions without whom ATLAS could not be
operated efficiently.
 
We acknowledge the support of
ANPCyT, Argentina;
YerPhI, Armenia;
ARC, Australia;
BMWFW and FWF, Austria;
ANAS, Azerbaijan;
SSTC, Belarus;
CNPq and FAPESP, Brazil;
NSERC, NRC and CFI, Canada;
CERN;
ANID, Chile;
CAS, MOST and NSFC, China;
Minciencias, Colombia;
MSMT CR, MPO CR and VSC CR, Czech Republic;
DNRF and DNSRC, Denmark;
IN2P3-CNRS and CEA-DRF/IRFU, France;
SRNSFG, Georgia;
BMBF, HGF and MPG, Germany;
GSRI, Greece;
RGC and Hong Kong SAR, China;
ISF and Benoziyo Center, Israel;
INFN, Italy;
MEXT and JSPS, Japan;
CNRST, Morocco;
NWO, Netherlands;
RCN, Norway;
MEiN, Poland;
FCT, Portugal;
MNE/IFA, Romania;
JINR;
MES of Russia and NRC KI, Russian Federation;
MESTD, Serbia;
MSSR, Slovakia;
ARRS and MIZ\v{S}, Slovenia;
DSI/NRF, South Africa;
MICINN, Spain;
SRC and Wallenberg Foundation, Sweden;
SERI, SNSF and Cantons of Bern and Geneva, Switzerland;
MOST, Taiwan;
TAEK, Turkey;
STFC, United Kingdom;
DOE and NSF, United States of America.
In addition, individual groups and members have received support from
BCKDF, CANARIE, Compute Canada and CRC, Canada;
COST, ERC, ERDF, Horizon 2020 and Marie Sk{\l}odowska-Curie Actions, European Union;
Investissements d'Avenir Labex, Investissements d'Avenir Idex and ANR, France;
DFG and AvH Foundation, Germany;
Herakleitos, Thales and Aristeia programmes co-financed by EU-ESF and the Greek NSRF, Greece;
BSF-NSF and GIF, Israel;
Norwegian Financial Mechanism 2014-2021, Norway;
NCN and NAWA, Poland;
La Caixa Banking Foundation, CERCA Programme Generalitat de Catalunya and PROMETEO and GenT Programmes Generalitat Valenciana, Spain;
G\"{o}ran Gustafssons Stiftelse, Sweden;
The Royal Society and Leverhulme Trust, United Kingdom.
 
The crucial computing support from all WLCG partners is acknowledged gratefully, in particular from CERN, the ATLAS Tier-1 facilities at TRIUMF (Canada), NDGF (Denmark, Norway, Sweden), CC-IN2P3 (France), KIT/GridKA (Germany), INFN-CNAF (Italy), NL-T1 (Netherlands), PIC (Spain), ASGC (Taiwan), RAL (UK) and BNL (USA), the Tier-2 facilities worldwide and large non-WLCG resource providers. Major contributors of computing resources are listed in Ref.~\cite{ATL-SOFT-PUB-2021-003}.
 
% End of text imported from the .//acknowledgements/Acknowledgements.tex input file

\printbibliography

@article{CMS:2019ybf,
    author = "Sirunyan, Albert M and others",
    collaboration = "CMS",
    title = "{Searches for physics beyond the standard model with the $M_\mathrm{T2}$ variable in hadronic final states with and without disappearing tracks in proton-proton collisions at $\sqrt{s}=$ 13 TeV}",
    eprint = "1909.03460",
    archivePrefix = "arXiv",
    primaryClass = "hep-ex",
    reportNumber = "CMS-SUS-19-005, CERN-EP-2019-180",
    doi = "10.1140/epjc/s10052-019-7493-x",
    journal = "Eur. Phys. J. C",
    volume = "80",
    number = "1",
    pages = "3",
    year = "2020"
}

@article{Debove:2010kf,
    author = "Debove, Jonathan and Fuks, Benjamin and Klasen, Michael",
    title = "{Threshold resummation for gaugino pair production at hadron colliders}",
    eprint = "1005.2909",
    archivePrefix = "arXiv",
    primaryClass = "hep-ph",
    reportNumber = "IPHC-PHENO-10-02, LPSC-10-050",
    doi = "10.1016/j.nuclphysb.2010.08.016",
    journal = "Nucl. Phys. B",
    volume = "842",
    pages = "51--85",
    year = "2011"
}

@article{PhysRevD.102.095021,
  title = {Higgsino and gaugino pair production at the LHC with $\mathrm{a}\mathrm{NNLO}+\mathrm{NNLL}$ precision},
  author = {Fiaschi, Juri and Klasen, Michael},
  journal = {Phys. Rev. D},
  volume = {102},
  issue = {9},
  pages = {095021},
  numpages = {14},
  year = {2020},
  month = {Nov},
  publisher = {American Physical Society},
  doi = {10.1103/PhysRevD.102.095021},
  url = {https://link.aps.org/doi/10.1103/PhysRevD.102.095021}
}

@article{Beenakker:1996ch,
    author = "Beenakker, W. and Hopker, R. and Spira, M. and Zerwas, P. M.",
    title = "{Squark and gluino production at hadron colliders}",
    eprint = "hep-ph/9610490",
    archivePrefix = "arXiv",
    reportNumber = "DESY-96-150, CERN-TH-96-215",
    doi = "10.1016/S0550-3213(97)80027-2",
    journal = "Nucl. Phys. B",
    volume = "492",
    pages = "51--103",
    year = "1997"
}

@techreport{Ellis:2691414,
      author        = "Ellis, Richard Keith and others",
      title         = "{Physics Briefing Book: Input for the European Strategy
                       for Particle Physics Update 2020}",
      address       = "Geneva",
      archivePrefix = "arXiv",
      eprint        = "1910.11775",
      type   = "CERN-ESU-004",
      month         = "Oct",
      year          = "2019",
      url           = "https://cds.cern.ch/record/2691414",
      note          = "254 p",
}

@article{LUCID2,
  author={G. Avoni and others},
  title={The new LUCID-2 detector for luminosity measurement and monitoring in ATLAS},
  journal={JINST},
  volume={13},
  number={07},
  pages={P07017},
  doi="10.1088/1748-0221/13/07/P07017",
  year={2018},
}

@article{PhysRevLett.81.34,
  title = {Phenomenology of Massive Vectorlike Doublet Leptons},
  author = {Thomas, Scott and Wells, James D.},
  journal = {Phys. Rev. Lett.},
  volume = {81},
  pages = {34--37},
  numpages = {0},
  year = {1998},
  month = {7},
  publisher = {American Physical Society},
  doi = {10.1103/PhysRevLett.81.34},
}

@article{Fukuda:2017jmk,
    author = "Fukuda, Hajime and Nagata, Natsumi and Otono, Hidetoshi and Shirai, Satoshi",
    title = "{Higgsino Dark Matter or Not: Role of Disappearing Track Searches at the LHC and Future Colliders}",
    eprint = "1703.09675",
    archivePrefix = "arXiv",
    primaryClass = "hep-ph",
    reportNumber = "IPMU-17-0050, UT-17-11, KYUSHU-RCAPP-2017-03",
    doi = "10.1016/j.physletb.2018.03.088",
    journal = "Phys. Lett. B",
    volume = "781",
    pages = "306--311",
    year = "2018"
}

@article{Papucci:2011wy,
    author = "Papucci, Michele and Ruderman, Joshua T. and Weiler, Andreas",
    title = "{Natural SUSY Endures}",
    eprint = "1110.6926",
    archivePrefix = "arXiv",
    primaryClass = "hep-ph",
    reportNumber = "DESY-11-193, CERN-PH-TH-265",
    doi = "10.1007/JHEP09(2012)035",
    journal = "JHEP",
    volume = "09",
    pages = "035",
    year = "2012"
}

@article{Ibe:2012sx,
    author = "Ibe, Masahiro and Matsumoto, Shigeki and Sato, Ryosuke",
    title = "{Mass Splitting between Charged and Neutral Winos at Two-Loop Level}",
    eprint = "1212.5989",
    archivePrefix = "arXiv",
    primaryClass = "hep-ph",
    reportNumber = "ICRR-REPORT-641-2012-30, IPMU-12-0231, UT-12-44",
    doi = "10.1016/j.physletb.2013.03.015",
    journal = "Phys. Lett. B",
    volume = "721",
    pages = "252--260",
    year = "2013"
}

@article{Paige:2003mg,
  author         = "Paige, Frank E. and Protopopescu, Serban D. and Baer,
  Howard and Tata, Xerxes",
  title          = "{ISAJET 7.69: A Monte Carlo Event Generator for $pp,~\bar{p}p$, and $e^+e^-$ Reactions}",
  year           = "2003",
  eprint         = "hep-ph/0312045",
  archivePrefix  = "arXiv",
  primaryClass   = "hep-ph",
}

@Booklet{ATL-PHYS-PUB-2009-002,
    author         = "{ATLAS Collaboration}",
    title          = "{The Expected Performance of the ATLAS Inner Detector}",
    howpublished   = "{ATL-PHYS-PUB-2009-002}",
    url            = "https://cds.cern.ch/record/1118445",
    year           = "2008",
}

@article{James:2004xla,
      author         = "James, Fred and Winkler, Matthias",
      title          = "{MINUIT User's Guide}",
      year           = "2004",
      url          = {http://seal.web.cern.ch/seal/documents/minuit/mnusersguide.pdf},
      SLACcitation   = "%%CITATION = INSPIRE-1258345;%%"
}

@article{Das:2016stf,
    author = "Das, Souvik",
    title = "{A simple alternative to the Crystal Ball function}",
    eprint = "1603.08591",
    archivePrefix = "arXiv",
    primaryClass = "hep-ex",
    month = "3",
    year = "2016"
}

@article{Golfand:1971iw,
      author         = {Golfand, Y.A.  and Likhtman, E.P.},
      title          = "{Extension of the Algebra of Poincare Group Generators
                        and Violation of P Invariance}",
      journal        = "JETP Lett.",
      volume         = "13",
      year           = "1971",
      pages          = "323",
      note           = "[Pisma Zh. Eksp. Teor. Fiz. \textbf{13} (1971) 452]",
      SLACcitation   = "%%CITATION = JTPLA,13,323;%%"
}

@article{Volkov:1973ix,
      author         = "Volkov, D.V. and Akulov, V.P.",
      title          = "{Is the neutrino a goldstone particle?}",
      journal        = "Phys. Lett. B",
      volume         = "46",
      year           = "1973",
      pages          = "109",
      doi            = "10.1016/0370-2693(73)90490-5",
      SLACcitation   = "%%CITATION = PHLTA,B46,109;%%"
}

@article{Wess:1974tw,
      author         = "Wess, J. and Zumino, B.",
      title          = "{Supergauge transformations in four dimensions}",
      journal        = "Nucl. Phys. B",
      volume         = "70",
      year           = "1974",
      pages          = "39",
      doi            = "10.1016/0550-3213(74)90355-1",
      SLACcitation   = "%%CITATION = NUPHA,B70,39;%%"
}

@article{Salam:1974ig,
      author         = "Salam, Abdus and Strathdee, J.",
      title          = "{Super-symmetry and non-Abelian gauges}",
      journal        = "Phys. Lett. B",
      volume         = "51",
      year           = "1974",
      pages          = "353",
      doi            = "10.1016/0370-2693(74)90226-3",
      reportNumber   = "IC/74/36",
      SLACcitation   = "%%CITATION = PHLTA,B51,353;%%"
}

@article{Wess:1974jb,
      author         = "Wess, J. and Zumino, B.",
      title          = "{Supergauge invariant extension of quantum electrodynamics}",
      journal        = "Nucl. Phys. B",
      volume         = "78",
      year           = "1974",
      pages          = "1",
      doi            = "10.1016/0550-3213(74)90112-6",
      reportNumber   = "CERN-TH-1857",
      SLACcitation   = "%%CITATION = NUPHA,B78,1;%%"
}

@article{Ferrara:1974pu,
      author         = "Ferrara, S. and Zumino, B.",
      title          = "{Supergauge invariant Yang-Mills theories}",
      journal        = "Nucl. Phys. B",
      volume         = "79",
      year           = "1974",
      pages          = "413",
      doi            = "10.1016/0550-3213(74)90559-8",
      reportNumber   = "CERN-TH-1866",
      SLACcitation   = "%%CITATION = NUPHA,B79,413;%%"
}

@article{Barbieri:1987fn,
      author         = "Barbieri, Riccardo and Giudice, G.F.",
      title          = "{Upper bounds on supersymmetric particle masses}",
      journal        = "Nucl. Phys. B",
      volume         = "306",
      year           = "1988",
      pages          = "63",
      doi            = "10.1016/0550-3213(88)90171-X",
      reportNumber   = "CERN-TH-4825/87",
      SLACcitation   = "%%CITATION = NUPHA,B306,63;%%"
}

@article{deCarlos:1993yy,
      author         = "de Carlos, B. and Casas, J.A.",
      title          = "{One-loop analysis of the electroweak breaking in supersymmetric models and the fine-tuning problem}",
      journal        = "Phys. Lett. B",
      volume         = "309",
      year           = "1993",
      pages          = "320",
      doi            = "10.1016/0370-2693(93)90940-J",
      eprint         = "hep-ph/9303291",
      archivePrefix  = "arXiv",
      reportNumber   = "CERN-TH-6835-93, IEM-FT-70-93",
      SLACcitation   = "%%CITATION = HEP-PH/9303291;%%"
}

@article{Giudice:1998xp,
      author         = "Giudice, Gian F. and Luty, Markus A. and Murayama, Hitoshi and Rattazzi, Riccardo",
      title          = "{Gaugino mass without singlets}",
      journal        = "JHEP",
      volume         = "12",
      year           = "1998",
      pages          = "027",
      doi            = "10.1088/1126-6708/1998/12/027",
      eprint         = "hep-ph/9810442",
      archivePrefix  = "arXiv",
      reportNumber   = "CERN-TH-98-337, LBNL-42419, LBL-42419, UCB-PTH-98-50,
                        UMD-PP-99-037",
      SLACcitation   = "%%CITATION = HEP-PH/9810442;%%"
}

@article{Randall:1998uk,
      author         = "Randall, Lisa and Sundrum, Raman",
      title          = "{Out of this world supersymmetry breaking}",
      journal        = "Nucl. Phys. B",
      volume         = "557",
      year           = "1999",
      pages          = "79",
      doi            = "10.1016/S0550-3213(99)00359-4",
      eprint         = "hep-th/9810155",
      archivePrefix  = "arXiv",
      reportNumber   = "MIT-CTP-2788, PUPT-1815, BUHEP-98-26",
      SLACcitation   = "%%CITATION = HEP-TH/9810155;%%"
}

@article{Fuks:2012qx,
     author         = "Fuks, Benjamin and Klasen, Michael and Lamprea, David R. and Rothering, Marcel",
     title          = "{Gaugino production in proton-proton collisions at a center-of-mass energy of \(8\,\text{TeV}\)}",
     journal        = "JHEP",
     volume         = "10",
     pages          = "081",
     doi            = "10.1007/JHEP10(2012)081",
     year           = "2012",
     eprint         = "1207.2159",
     archivePrefix  = "arXiv",
     primaryClass   = "hep-ph",
     reportNumber   = "IPHC-PHENO-12-07, MS-TP-12-05",
     SLACcitation   = "%%CITATION = ARXIV:1207.2159;%%",
}

@article{Fuks:2013vua,
     author         = "Fuks, Benjamin and Klasen, Michael and Lamprea, David R. and Rothering, Marcel",
     title          = "{Precision predictions for electroweak superpartner production at hadron colliders with \textsc{resummino}}",
     journal        = "Eur. Phys. J. C",
     volume         = "73",
     pages          = "2480",
     doi            = "10.1140/epjc/s10052-013-2480-0",
     year           = "2013",
     eprint         = "1304.0790",
     archivePrefix  = "arXiv",
     primaryClass   = "hep-ph",
     reportNumber   = "CERN-PH-TH-2013-064, IPHC-PHENO-13-02, MS-TP-13-06",
     SLACcitation   = "%%CITATION = ARXIV:1304.0790;%%",
}

@Article{Cacciari:2008gp,
     author    = "Cacciari, Matteo and Salam, Gavin P. and Soyez, Gregory",
     title     = "{The anti-\(k_{t}\) jet clustering algorithm}",
     journal   = "JHEP",
     volume    = "04",
     year      = "2008",
     pages     = "063",
     eprint    = "0802.1189",
     archivePrefix = "arXiv",
     primaryClass  =  "hep-ph",
     doi       = "10.1088/1126-6708/2008/04/063",
     SLACcitation  = "%%CITATION = 0802.1189;%%"
}

@Article{Fastjet,
      author         = "Cacciari, Matteo and Salam, Gavin P. and Soyez, Gregory",
      title          = "{FastJet user manual}",
      journal        = "Eur. Phys. J. C",
      volume         = "72",
      year           = "2012",
      pages          = "1896",
      doi            = "10.1140/epjc/s10052-012-1896-2",
      eprint         = "1111.6097",
      archivePrefix  = "arXiv",
      primaryClass   = "hep-ph",
      reportNumber   = "CERN-PH-TH-2011-297",
      SLACcitation   = "%%CITATION = ARXIV:1111.6097;%%"
}

@Article{Butterworth:2015oua,
      author         = "Butterworth, Jon and others",
      title          = "{PDF4LHC recommendations for LHC Run II}",
      journal        = "J. Phys. G",
      volume         = "43",
      year           = "2016",
      pages          = "023001",
      doi            = "10.1088/0954-3899/43/2/023001",
      eprint         = "1510.03865",
      archivePrefix  = "arXiv",
      primaryClass   = "hep-ph",
      reportNumber   = "OUTP-15-17P, SMU-HEP-15-12, TIF-UNIMI-2015-14,
                        LCTS-2015-27, CERN-PH-TH-2015-249",
      SLACcitation   = "%%CITATION = ARXIV:1510.03865;%%"
}

@Article{Pumplin:2002vw,
     author    = "Pumplin, J. and others",
     title     = "{New Generation of Parton Distributions with Uncertainties from Global QCD Analysis}",
     journal   = "JHEP",
     volume    = "07",
     year      = "2002",
     pages     = "012",
     doi       = "10.1088/1126-6708/2002/07/012",
     eprint    = "hep-ph/0201195",
     archivePrefix = "arXiv",
     SLACcitation  = "%%CITATION = HEP-PH/0201195;%%"
}

@article{Martin:2009iq,
      author         = "Martin, A. D. and Stirling, W. J. and Thorne, R. S. and Watt, G.",
      title          = "{Parton distributions for the LHC}",
      journal        = "Eur. Phys. J. C",
      volume         = "63",
      year           = "2009",
      pages          = "189",
      doi            = "10.1140/epjc/s10052-009-1072-5",
      eprint         = "0901.0002",
      archivePrefix  = "arXiv",
      primaryClass   = "hep-ph",
      reportNumber   = "IPPP-08-95, DCPT-08-190, CAVENDISH-HEP-08-16",
      SLACcitation   = "%%CITATION = ARXIV:0901.0002;%%"
}

@article{Ball:2012cx,
      author         = "Ball, Richard D. and others",
      title          = "{Parton distributions with LHC data}",
      journal        = "Nucl. Phys. B",
      volume         = "867",
      year           = "2013",
      pages          = "244",
      doi            = "10.1016/j.nuclphysb.2012.10.003",
      eprint         = "1207.1303",
      archivePrefix  = "arXiv",
      primaryClass   = "hep-ph",
      reportNumber   = "EDINBURGH-2012-08, IFUM-FT-997, FR-PHENO-2012-014,
                        RWTH-TTK-12-25, CERN-PH-TH-2012-037, SFB-CPP-12-47\,
                        --CERN-PH-TH-2012-037",
      SLACcitation   = "%%CITATION = ARXIV:1207.1303;%%"
}

@article{Ball:2014uwa,
      author         = "Ball, Richard D. and others",
      title          = "{Parton distributions for the LHC run II}",
      collaboration  = "NNPDF",
      journal        = "JHEP",
      volume         = "04",
      year           = "2015",
      pages          = "040",
      doi            = "10.1007/JHEP04(2015)040",
      eprint         = "1410.8849",
      archivePrefix  = "arXiv",
      primaryClass   = "hep-ph",
      reportNumber   = "EDINBURGH-2014-15, IFUM-1034-FT, CERN-PH-TH-2013-253,
                        OUTP-14-11P, CAVENDISH-HEP-14-11",
      SLACcitation   = "%%CITATION = ARXIV:1410.8849;%%"
}

@Article{Sjostrand:2007gs,
	author    = "Sj{\"o}strand, T. and Mrenna, S. and Skands, P.",
	title     = "{A brief introduction to PYTHIA 8.1}",
	journal   = "Comput. Phys. Commun.",
	volume    = "178",
	year      = "2008",
	pages     = "852-867",
	eprint    = "0710.3820",
	archivePrefix = "arXiv",
	primaryClass  =  "hep-ph",
	doi       = "10.1016/j.cpc.2008.01.036",
	SLACcitation  = "%%CITATION = 0710.3820;%%"
}

@article{Sjostrand:2014zea,
      author         = "Sj{\"o}strand, Torbj{\"o}rn and Ask, Stefan and Christiansen,
                        Jesper R. and Corke, Richard and Desai, Nishita and Ilten,
                        Philip and Mrenna, Stephen and Prestel, Stefan and
                        Rasmussen, Christine O. and Skands, Peter Z.",
      title          = "{An introduction to PYTHIA 8.2}",
      journal        = "Comput. Phys. Commun.",
      volume         = "191",
      year           = "2015",
      pages          = "159",
      doi            = "10.1016/j.cpc.2015.01.024",
      eprint         = "1410.3012",
      archivePrefix  = "arXiv",
      primaryClass   = "hep-ph",
      reportNumber   = "LU-TP-14-36, MCNET-14-22, CERN-PH-TH-2014-190,
                        FERMILAB-PUB-14-316-CD, DESY-14-178, SLAC-PUB-16122,
                        --FERMILAB-PUB-14-316-CD",
      SLACcitation   = "%%CITATION = ARXIV:1410.3012;%%"
}

@article{Lonnblad:2011xx,
      author         = "L{\"o}nnblad, Leif and Prestel, Stefan",
      title          = "{Matching tree-level matrix elements with interleaved
                        showers}",
      journal        = "JHEP",
      volume         = "03",
      year           = "2012",
      pages          = "019",
      doi            = "10.1007/JHEP03(2012)019",
      eprint         = "1109.4829",
      archivePrefix  = "arXiv",
      primaryClass   = "hep-ph",
      SLACcitation   = "%%CITATION = ARXIV:1109.4829;%%"
}

@Article{Lange:2001uf,
      author         = "Lange, D. J.",
      title          = "{The EvtGen particle decay simulation package}",
      booktitle      = "{Proceedings, 7th International Conference on B physics
                        at hadron machines (BEAUTY 2000)}",
      journal        = "Nucl. Instrum. Meth. A",
      volume         = "462",
      year           = "2001",
      pages          = "152",
      doi            = "10.1016/S0168-9002(01)00089-4",
      SLACcitation   = "%%CITATION = NUIMA,A462,152;%%"
}

@Article{Alwall:2014hca,
      author         = "Alwall, J. and Frederix, R. and Frixione, S. and Hirschi,
                        V. and Maltoni, F. and Mattelaer, O. and Shao, H. -S. and
                        Stelzer, T. and Torrielli, P. and Zaro, M.",
      title          = "{The automated computation of tree-level and
                        next-to-leading order differential cross sections, and
                        their matching to parton shower simulations}",
      journal        = "JHEP",
      volume         = "07",
      year           = "2014",
      pages          = "079",
      doi            = "10.1007/JHEP07(2014)079",
      eprint         = "1405.0301",
      archivePrefix  = "arXiv",
      primaryClass   = "hep-ph",
      reportNumber   = "CERN-PH-TH-2014-064, CP3-14-18, LPN14-066, MCNET-14-09,
                        ZU-TH-14-14",
      SLACcitation   = "%%CITATION = ARXIV:1405.0301;%%"
}

@Article{Nason:2004rx,
      author         = "Nason, Paolo",
      title          = "{A new method for combining NLO QCD with shower Monte Carlo algorithms}",
      journal        = "JHEP",
      volume         = "11",
      pages          = "040",
      doi            = "10.1088/1126-6708/2004/11/040",
      year           = "2004",
      eprint         = "hep-ph/0409146",
      archivePrefix  = "arXiv",
}

@Article{Frixione:2007vw,
      author         = "Frixione, Stefano and Nason, Paolo and Oleari, Carlo",
      title          = "{Matching NLO QCD computations with parton shower
                        simulations: the POWHEG method}",
      journal        = "JHEP",
      volume         = "11",
      pages          = "070",
      doi            = "10.1088/1126-6708/2007/11/070",
      year           = "2007",
      eprint         = "0709.2092",
      archivePrefix  = "arXiv",
      primaryClass   = "hep-ph",
}

@Article{Alioli:2010xd,
      author         = "Alioli, Simone and Nason, Paolo and Oleari, Carlo and Re,
                        Emanuele",
      title          = "{A general framework for implementing NLO calculations in
                        shower Monte Carlo programs: the POWHEG BOX}",
      journal        = "JHEP",
      volume         = "06",
      pages          = "043",
      doi            = "10.1007/JHEP06(2010)043",
      year           = "2010",
      eprint         = "1002.2581",
      archivePrefix  = "arXiv",
      primaryClass   = "hep-ph",
}

@article{Frixione:2007nw,
      author         = "Frixione, Stefano and Nason, Paolo and Ridolfi, Giovanni",
      title          = "{A positive-weight next-to-leading-order Monte Carlo for
                        heavy flavour hadroproduction}",
      journal        = "JHEP",
      volume         = "09",
      pages          = "126",
      doi            = "10.1088/1126-6708/2007/09/126",
      year           = "2007",
      eprint         = "0707.3088",
      archivePrefix  = "arXiv",
      primaryClass   = "hep-ph",
      SLACcitation   = "%%CITATION = ARXIV:0707.3088;%%",
}

@Article{Hoeche:2009rj,
      author         = "H{\"o}che, Stefan and Krauss, Frank and Schumann, Steffen and Siegert, Frank",
      title          = "{QCD matrix elements and truncated showers}",
      journal        = "JHEP",
      volume         = "05",
      pages          = "053",
      doi            = "10.1088/1126-6708/2009/05/053",
      year           = "2009",
      eprint         = "0903.1219",
      archivePrefix  = "arXiv",
      primaryClass   = "hep-ph",
}

@Article{Gleisberg:2008fv,
      author         = "Gleisberg, Tanju and H{\"o}che, Stefan",
      title          = "{Comix, a new matrix element generator}",
      journal        = "JHEP",
      volume         = "12",
      pages          = "039",
      doi            = "10.1088/1126-6708/2008/12/039",
      year           = "2008",
      eprint         = "0808.3674",
      archivePrefix  = "arXiv",
      primaryClass   = "hep-ph",
}

@Article{Schumann:2007mg,
      author         = "Schumann, Steffen and Krauss, Frank",
      title          = "{A parton shower algorithm based on Catani--Seymour dipole factorisation}",
      journal        = "JHEP",
      volume         = "03",
      pages          = "038",
      doi            = "10.1088/1126-6708/2008/03/038",
      year           = "2008",
      eprint         = "0709.1027",
      archivePrefix  = "arXiv",
      primaryClass   = "hep-ph",
}

@article{Hoeche:2012yf,
      author         = "H{\"o}che, Stefan and Krauss, Frank and Sch{\"o}nherr, Marek and
                        Siegert, Frank",
      title          = "{QCD matrix elements + parton showers. The NLO case}",
      journal        = "JHEP",
      volume         = "04",
      year           = "2013",
      pages          = "027",
      doi            = "10.1007/JHEP04(2013)027",
      eprint         = "1207.5030",
      archivePrefix  = "arXiv",
      primaryClass   = "hep-ph",
      reportNumber   = "SLAC-PUB-15191, IPPP-12-52, DCPT-12-104, LPN12-081,
                        FR-PHENO-2012-017, MCNET-12-09, --FR-PHENO-2012-017",
      SLACcitation   = "%%CITATION = ARXIV:1207.5030;%%"
}

@article{Bothmann:2019yzt,
      author = "Bothmann, Enrico and others",
      title = "{Event generation with Sherpa 2.2}",
      journal = "SciPost Phys.",
      volume = "7",
      year = "2019",
      number = "3",
      pages = "034",
      doi = "10.21468/SciPostPhys.7.3.034",
      reportNumber = "FERMILAB-PUB-19-218-T, SLAC-PUB-17433, IPPP/19/42, MCNET-19-11",
      eprint = "1905.09127",
      archivePrefix = "arXiv",
      primaryClass = "hep-ph",
}

@article{Hoeche:2011fd,
      author         = "H{\"o}che, Stefan and Krauss, Frank and Sch{\"o}nherr, Marek and
                        Siegert, Frank",
      title          = "{A critical appraisal of NLO+PS matching methods}",
      journal        = "JHEP",
      volume         = "09",
      year           = "2012",
      pages          = "049",
      doi            = "10.1007/JHEP09(2012)049",
      eprint         = "1111.1220",
      archivePrefix  = "arXiv",
      primaryClass   = "hep-ph",
      reportNumber   = "SLAC-PUB-14661, IPPP-11-67, DCPT-11-134, LPN11-58,
                        FR-PHENO-2011-019, MCNET-11-24",
      SLACcitation   = "%%CITATION = ARXIV:1111.1220;%%"
}

@article{Catani:2001cc,
      author         = "Catani, S. and Krauss, F. and Webber, B. R. and Kuhn, R.",
      title          = "{QCD Matrix Elements + Parton Showers}",
      journal        = "JHEP",
      volume         = "11",
      year           = "2001",
      pages          = "063",
      doi            = "10.1088/1126-6708/2001/11/063",
      eprint         = "hep-ph/0109231",
      archivePrefix  = "arXiv",
      reportNumber   = "CERN-TH-2000-367, CAVENDISH-HEP-00-03",
      SLACcitation   = "%%CITATION = HEP-PH/0109231;%%"
}

@article{Cascioli:2011va,
      author         = "Cascioli, Fabio and Maierh{\"o}fer, Philipp and Pozzorini, Stefano",
      title          = "{Scattering Amplitudes with Open Loops}",
      journal        = "Phys. Rev. Lett.",
      volume         = "108",
      year           = "2012",
      pages          = "111601",
      doi            = "10.1103/PhysRevLett.108.111601",
      eprint         = "1111.5206",
      archivePrefix  = "arXiv",
      primaryClass   = "hep-ph",
      reportNumber   = "ZU-TH-23-11, LPN11-66",
      SLACcitation   = "%%CITATION = ARXIV:1111.5206;%%"
}

@article{Buccioni:2019sur,
    author = "Buccioni, Federico and Lang, Jean-Nicolas and Lindert, Jonas M. and Maierh{\"o}fer, Philipp and Pozzorini, Stefano and Zhang, Hantian and Zoller, Max F.",
    title = "{OpenLoops 2}",
    eprint = "1907.13071",
    archivePrefix = "arXiv",
    primaryClass = "hep-ph",
    reportNumber = "IPPP/19/62, FR-PHENO-2019-12, PSI-PR-19-15, ZU-TH 37/19",
    doi = "10.1140/epjc/s10052-019-7306-2",
    journal = "Eur. Phys. J. C",
    volume = "79",
    number = "10",
    pages = "866",
    year = "2019"
}

@article{Denner:2016kdg,
      author         = "Denner, Ansgar and Dittmaier, Stefan and Hofer, Lars",
      title          = "{\textsc{Collier}: A fortran-based complex one-loop library in
                        extended regularizations}",
      journal        = "Comput. Phys. Commun.",
      volume         = "212",
      year           = "2017",
      pages          = "220-238",
      doi            = "10.1016/j.cpc.2016.10.013",
      eprint         = "1604.06792",
      archivePrefix  = "arXiv",
      primaryClass   = "hep-ph",
      reportNumber   = "FR-PHENO-2016-003, ICCUB-16-016",
      SLACcitation   = "%%CITATION = ARXIV:1604.06792;%%"
}

@article{Agostinelli:2002hh,
      author         = "{GEANT4 Collaboration} and Agostinelli, S. and others",
      title          = "{\textsc{Geant4} -- a simulation toolkit}",
      journal        = "Nucl. Instrum. Meth. A",
      volume         = "506",
      year           = "2003",
      pages          = "250",
      doi            = "10.1016/S0168-9002(03)01368-8",
      reportNumber   = "SLAC-PUB-9350, FERMILAB-PUB-03-339",
      SLACcitation   = "%%CITATION = NUIMA,A506,250;%%"
}

@Article{Read:2002hq,
    Author = {Read, Alexander L.},
    Journal = {J. Phys. G},
    Pages = {2693},
    doi = "10.1088/0954-3899/28/10/313",
    Title = "{Presentation of search results: the \(CL_S\) technique}",
    Volume = {28},
    Year = {2002}
}

@Booklet{RooFit,
      author         = "Verkerke, Wouter and Kirkby, David",
      title          = "{The RooFit toolkit for data modeling}",
      year           = "2003",
      eprint         = "physics/0306116",
      archivePrefix  = "arXiv",
      primaryClass   = "physics.data-an",
}

@Article{PERF-2007-01,
    author         = "{ATLAS Collaboration}",
    title          = "{The ATLAS Experiment at the CERN Large Hadron Collider}",
    journal        = "JINST",
    volume         = "3",
    year           = "2008",
    pages          = "S08003",
    doi            = "10.1088/1748-0221/3/08/S08003",
    primaryClass   = "hep-ex",
}

@Article{SOFT-2010-01,
    author         = "{ATLAS Collaboration}",
    title          = "{The ATLAS Simulation Infrastructure}",
    journal        = "Eur. Phys. J. C",
    volume         = "70",
    year           = "2010",
    pages          = "823",
    doi            = "10.1140/epjc/s10052-010-1429-9",
    eprint         = "1005.4568",
    archivePrefix  = "arXiv",
    primaryClass   = "physics.ins-det",
}

@Article{SUSY-2013-01,
    author         = "{ATLAS Collaboration}",
    title          = "{Search for charginos nearly mass degenerate with the lightest neutralino based on a disappearing-track signature in \(pp\) collisions at \(\sqrt{s} =8\,\text{TeV}\) with the ATLAS detector}",
    journal        = "Phys. Rev. D",
    volume         = "88",
    year           = "2013",
    pages          = "112006",
    doi            = "10.1103/PhysRevD.88.112006",
    reportNumber   = "CERN-PH-EP-2013-155",
    eprint         = "1310.3675",
    archivePrefix  = "arXiv",
    primaryClass   = "hep-ex",
}

@Article{PERF-2014-03,
    author         = "{ATLAS Collaboration}",
    title          = "{Performance of pile-up mitigation techniques for jets in \(pp\) collisions at \(\sqrt{s} = 8\,\text{TeV}\) using the ATLAS detector}",
    journal        = "Eur. Phys. J. C",
    volume         = "76",
    year           = "2016",
    pages          = "581",
    doi            = "10.1140/epjc/s10052-016-4395-z",
    reportNumber   = "CERN-PH-EP-2015-206",
    eprint         = "1510.03823",
    archivePrefix  = "arXiv",
    primaryClass   = "hep-ex",
}

@Article{PERF-2016-07,
    author         = "{ATLAS Collaboration}",
    title          = "{Performance of missing transverse momentum reconstruction with the ATLAS detector using proton--proton collisions at \(\sqrt{s} = 13\,\text{TeV}\)}",
    journal        = "Eur. Phys. J. C",
    volume         = "78",
    year           = "2018",
    pages          = "903",
    doi            = "10.1140/epjc/s10052-018-6288-9",
    reportNumber   = "CERN-EP-2017-274",
    eprint         = "1802.08168",
    archivePrefix  = "arXiv",
    primaryClass   = "hep-ex",
}

@Article{SUSY-2016-06,
    author         = "{ATLAS Collaboration}",
    title          = "{Search for long-lived charginos based on a disappearing-track signature in \(pp\) collisions at \(\sqrt{s} = 13\,\text{TeV}\) with the ATLAS detector}",
    journal        = "JHEP",
    volume         = "06",
    year           = "2018",
    pages          = "022",
    doi            = "10.1007/JHEP06(2018)022",
    reportNumber   = "CERN-EP-2017-179",
    eprint         = "1712.02118",
    archivePrefix  = "arXiv",
    primaryClass   = "hep-ex",
}

@Article{TRIG-2016-01,
    author         = "{ATLAS Collaboration}",
    title          = "{Performance of the ATLAS trigger system in 2015}",
    journal        = "Eur. Phys. J. C",
    volume         = "77",
    year           = "2017",
    pages          = "317",
    doi            = "10.1140/epjc/s10052-017-4852-3",
    reportNumber   = "CERN-EP-2016-241",
    eprint         = "1611.09661",
    archivePrefix  = "arXiv",
    primaryClass   = "hep-ex",
}

@Article{PERF-2017-01,
    author         = "{ATLAS Collaboration}",
    title          = "{Electron reconstruction and identification in the ATLAS experiment using the 2015 and 2016 LHC proton--proton collision data at \(\sqrt{s} = 13\,\text{TeV}\)}",
    journal        = "Eur. Phys. J. C",
    volume         = "79",
    year           = "2019",
    pages          = "639",
    doi            = "10.1140/epjc/s10052-019-7140-6",
    reportNumber   = "CERN-EP-2018-273",
    eprint         = "1902.04655",
    archivePrefix  = "arXiv",
    primaryClass   = "hep-ex",
}

@Article{DAPR-2018-01,
    author         = "{ATLAS Collaboration}",
    title          = "{ATLAS data quality operations and performance for 2015--2018 data-taking}",
    journal        = "JINST",
    volume         = "15",
    year           = "2020",
    pages          = "P04003",
    doi            = "10.1088/1748-0221/15/04/P04003",
    reportNumber   = "CERN-EP-2019-207",
    eprint         = "1911.04632",
    archivePrefix  = "arXiv",
    primaryClass   = "physics.ins-det",
}

@Article{EGAM-2018-01,
    author         = "{ATLAS Collaboration}",
    title          = "{Electron and photon performance measurements with the ATLAS detector using the 2015--2017 LHC proton--proton collision data}",
    journal        = "JINST",
    volume         = "14",
    year           = "2019",
    pages          = "P12006",
    doi            = "10.1088/1748-0221/14/12/P12006",
    reportNumber   = "CERN-EP-2019-145",
    eprint         = "1908.00005",
    archivePrefix  = "arXiv",
    primaryClass   = "hep-ex",
}

@Article{JETM-2018-05,
    author         = "{ATLAS Collaboration}",
    title          = "{Jet energy scale and resolution measured in proton--proton collisions at \(\sqrt{s} = 13\,\text{TeV}\) with the ATLAS detector}",
    journal        = "Eur. Phys. J. C",
    volume         = "81",
    year           = "2020",
    pages          = "689",
    doi            = "10.1140/epjc/s10052-021-09402-3",
    reportNumber   = "CERN-EP-2020-083",
    eprint         = "2007.02645",
    archivePrefix  = "arXiv",
    primaryClass   = "hep-ex",
}

@Article{MUON-2018-03,
    author         = "{ATLAS Collaboration}",
    title          = "{Muon reconstruction and identification efficiency in ATLAS using the full Run~2 \(pp\) collision data set at \(\sqrt{s} = 13\,\text{TeV}\)}",
    journal        = "Eur. Phys. J. C",
    volume         = "81",
    year           = "2020",
    pages          = "578",
    doi            = "10.1140/epjc/s10052-021-09233-2",
    reportNumber   = "CERN Preprint ID: CERN-EP-2020-199",
    eprint         = "2012.00578",
    archivePrefix  = "arXiv",
    primaryClass   = "hep-ex",
}

@Article{TRIG-2018-01,
    author         = "{ATLAS Collaboration}",
    title          = "{Performance of the ATLAS muon triggers in Run~2}",
    journal        = "JINST",
    volume         = "15",
    year           = "2020",
    pages          = "P09015",
    doi            = "10.1088/1748-0221/15/09/p09015",
    reportNumber   = "CERN-EP-2020-031",
    eprint         = "2004.13447",
    archivePrefix  = "arXiv",
    primaryClass   = "hep-ex",
}

@Article{TRIG-2018-05,
    author         = "{ATLAS Collaboration}",
    title          = "{Performance of electron and photon triggers in ATLAS during LHC Run~2}",
    journal        = "Eur. Phys. J. C",
    volume         = "80",
    year           = "2020",
    pages          = "47",
    doi            = "10.1140/epjc/s10052-019-7500-2",
    reportNumber   = "CERN-EP-2019-169",
    eprint         = "1909.00761",
    archivePrefix  = "arXiv",
    primaryClass   = "hep-ex",
}

@Article{TRIG-2019-01,
    author         = "{ATLAS Collaboration}",
    title          = "{Performance of the missing transverse momentum triggers for the ATLAS detector during Run-2 data taking}",
    journal        = "JHEP",
    volume         = "08",
    year           = "2020",
    pages          = "080",
    doi            = "10.1007/JHEP08(2020)080",
    reportNumber   = "CERN-EP-2020-050",
    eprint         = "2005.09554",
    archivePrefix  = "arXiv",
    primaryClass   = "hep-ex",
}

@Booklet{ATL-SOFT-PUB-2021-001,
    author         = "{ATLAS Collaboration}",
    title          = "{The ATLAS Collaboration Software and Firmware}",
    howpublished   = "{ATL-SOFT-PUB-2021-001}",
    url            = "https://cds.cern.ch/record/2767187",
    year           = "2021",
}

@Booklet{ATL-SOFT-PUB-2021-003,
    author         = "{ATLAS Collaboration}",
    title          = "{ATLAS Computing Acknowledgements}",
    howpublished   = "{ATL-SOFT-PUB-2021-003}",
    url            = "https://cds.cern.ch/record/2776662",
}

@Report{ATLAS-TDR-2010-19,
    author         = "{ATLAS Collaboration}",
    title          = "{ATLAS Insertable B-Layer Technical Design Report}",
    type           = "ATLAS-TDR-19; CERN-LHCC-2010-013",
    year           = "2010",
    url            = "https://cds.cern.ch/record/1291633",
    related        = "ATLAS-TDR-2010-19-addm",
    relatedstring  = "Addendum:",
}

@Article{CMS-EXO-19-010,
    author         = "{CMS Collaboration}",
    title          = "{Search for disappearing tracks in proton--proton collisions at \(\sqrt{s} = 13\,\text{TeV}\)}",
    journal        = "Phys. Lett. B",
    volume         = "806",
    year           = "2020",
    pages          = "135502",
    doi            = "10.1016/j.physletb.2020.135502",
    reportNumber   = "CERN-EP-2020-043",
    eprint         = "2004.05153",
    archivePrefix  = "arXiv",
    primaryClass   = "hep-ex",
}

@Booklet{ATLAS-CONF-2018-023,
    author         = "{ATLAS Collaboration}",
    title          = "{\(E_{\text{T}}^{\text{miss}}\) performance in the ATLAS detector using 2015--2016 LHC \(pp\) collisions}",
    howpublished   = "{ATLAS-CONF-2018-023}",
    url            = "https://cds.cern.ch/record/2625233",
    year           = "2018",
}

@Booklet{ATLAS-CONF-2019-021,
    author         = "{ATLAS Collaboration}",
    title          = "{Luminosity determination in \(pp\) collisions at \(\sqrt{s} = 13\,\text{TeV}\) using the ATLAS detector at the LHC}",
    howpublished   = "{ATLAS-CONF-2019-021}",
    url            = "https://cds.cern.ch/record/2677054",
    year           = "2019",
}

@Booklet{ATL-PHYS-PUB-2014-021,
    author         = "{ATLAS Collaboration}",
    title          = "{ATLAS Pythia~8 tunes to \(7~\text{TeV}\) data}",
    howpublished   = "{ATL-PHYS-PUB-2014-021}",
    url            = "https://cds.cern.ch/record/1966419",
    year           = "2014",
}

@Booklet{ATL-PHYS-PUB-2015-018,
    author         = "{ATLAS Collaboration}",
    title          = "{Track Reconstruction Performance of the ATLAS Inner Detector at  \(\sqrt{s} = 13~\text{TeV}\)}",
    howpublished   = "{ATL-PHYS-PUB-2015-018}",
    url            = "https://cds.cern.ch/record/2037683",
    year           = "2015",
}

@Booklet{ATL-PHYS-PUB-2016-020,
    author         = "{ATLAS Collaboration}",
    title          = "{Studies on top-quark Monte Carlo modelling for Top2016}",
    howpublished   = "{ATL-PHYS-PUB-2016-020}",
    url            = "https://cds.cern.ch/record/2216168",
    year           = "2016",
}

@Booklet{ATL-PHYS-PUB-2017-019,
    author         = "{ATLAS Collaboration}",
    title          = "{Search for direct pair production of higgsinos by reinterpretation of the disappearing track analysis with \(36.1~\text{fb}^{-1}\) of \(\sqrt{s} = 13~\text{TeV}\) data collected with the ATLAS experiment}",
    howpublished   = "{ATL-PHYS-PUB-2017-019}",
    url            = "https://cds.cern.ch/record/2297480",
    year           = "2017",
}

\clearpage
% ATLAS Collaboration author list
% Reference date of SUSY-2018-19 is 2021-03-03
% Author list last updated on date 26-JUL-22
% Data extracted on 26-Jul-2022 for paper reference SUSY-2018-19
% at 10:26pm
 
\begin{flushleft}
\hypersetup{urlcolor=black}
{\Large The ATLAS Collaboration}

\bigskip

\AtlasOrcid[0000-0002-6665-4934]{G.~Aad}$^\textrm{\scriptsize 99}$,    
\AtlasOrcid[0000-0002-5888-2734]{B.~Abbott}$^\textrm{\scriptsize 124}$,    
\AtlasOrcid[0000-0002-7248-3203]{D.C.~Abbott}$^\textrm{\scriptsize 100}$,    
\AtlasOrcid[0000-0002-2788-3822]{A.~Abed~Abud}$^\textrm{\scriptsize 34}$,    
\AtlasOrcid[0000-0002-1002-1652]{K.~Abeling}$^\textrm{\scriptsize 51}$,    
\AtlasOrcid[0000-0002-2987-4006]{D.K.~Abhayasinghe}$^\textrm{\scriptsize 91}$,    
\AtlasOrcid[0000-0002-8496-9294]{S.H.~Abidi}$^\textrm{\scriptsize 27}$,    
\AtlasOrcid[0000-0001-5329-6640]{H.~Abramowicz}$^\textrm{\scriptsize 157}$,    
\AtlasOrcid[0000-0002-1599-2896]{H.~Abreu}$^\textrm{\scriptsize 156}$,    
\AtlasOrcid[0000-0003-0403-3697]{Y.~Abulaiti}$^\textrm{\scriptsize 5}$
% ...
\bigskip
\\

% ...
$^{5}$High Energy Physics Division, Argonne National Laboratory, Argonne IL; United States of America.\\
% ...
$^{27}$Physics Department, Brookhaven National Laboratory, Upton NY; United States of America.\\
% ...
$^{34}$CERN, Geneva; Switzerland.\\
% ...
$^{51}$II. Physikalisches Institut, Georg-August-Universit\"{a}t G\"ottingen, G\"ottingen; Germany.\\
% ...
$^{91}$Department of Physics, Royal Holloway University of London, Egham; United Kingdom.\\
% ...
$^{99}$CPPM, Aix-Marseille Universit\'e, CNRS/IN2P3, Marseille; France.\\
$^{100}$Department of Physics, University of Massachusetts, Amherst MA; United States of America.\\
% ...
$^{124}$Homer L. Dodge Department of Physics and Astronomy, University of Oklahoma, Norman OK; United States of America.\\
% ...
$^{156}$Department of Physics, Technion, Israel Institute of Technology, Haifa; Israel.\\
$^{157}$Raymond and Beverly Sackler School of Physics and Astronomy, Tel Aviv University, Tel Aviv; Israel.\\
% ...

\end{flushleft}

% Created with Glance <Atlas.Glance@cern.ch>

\end{document}